%% file: mainV3.tex
\documentclass[
    twocolumn,
	prd,
	amssymb,
	preprintnumbers,superscriptaddress,
	nofootinbib]{revtex4-1}

\pdfoutput=1

\usepackage{graphicx}
\usepackage{enumitem}
\usepackage{latexsym}
\usepackage{amsfonts}
\usepackage{amssymb}
\usepackage{xcolor}
\usepackage[export]{adjustbox}
\usepackage{amsmath}
\usepackage[thinlines]{easytable}
\usepackage{slashed}
\usepackage{dcolumn}
\usepackage{verbatim}
\usepackage{float}
\usepackage{multirow}
\usepackage{xspace}
\usepackage[normalem]{ulem}
\usepackage[
pdfauthor={Djuna Croon}]{hyperref}
\usepackage{tabularx}

\usepackage[version=4]{mhchem}

\input{universalnewcommands.tex}

\renewcommand{\paragraph}[1]{~\\ \noindent{\bf \emph{#1} --}}
\renewcommand{\subparagraph}[1]{~\\ \noindent{ \emph{#1} --}}

\allowdisplaybreaks

\begin{document}

\title{Dark Matter (H)eats Young Planets}
\author{Djuna Croon} 
\thanks{\href{mailto:djuna.l.croon@durham.ac.uk}{djuna.l.croon@durham.ac.uk}; {\scriptsize ORCID}: \href{https://orcid.org/0000-0003-3359-3706}{0000-0003-3359-3706}}
\affiliation{Institute for Particle Physics Phenomenology, Department of Physics, Durham University, Durham DH1 3LE, U.K.}

\author{Juri Smirnov} 
\thanks{\href{mailto:juri.smirnov@liverpool.ac.uk}{juri.smirnov@liverpool.ac.uk}; {\scriptsize ORCID}: \href{http://orcid.org/0000-0002-3082-0929}{0000-0002-3082-0929}}
\affiliation{Department of Mathematical Sciences, University of Liverpool,
Liverpool, L69 7ZL, United Kingdom}

\date{\today}

\begin{abstract}
We study the effect of dark matter annihilation on the formation of Jovian planets. We show that dark matter heat injections can slow or halt Kelvin-Helmholtz contraction, preventing the accretion of hydrogen and helium onto the solid core. The existence of Jupiter in our solar system can therefore be used to infer constraints on dark matter with relatively strong interaction cross sections. 
We derive novel constraints on the cross section for both spin-dependent and spin-independent dark matter.
We highlight the possibility of a positive detection using future observations by JWST, which could reveal strongly varying planet morphologies close to our Galactic Center.
\end{abstract}

\preprint{IPPP/23/43}
\preprint{LTH-1349}

\maketitle

Our technical capabilities to detect planets outside our solar system have developed dramatically. From the first indirect evidence for an exoplanet~\cite{1995Natur.378..355M}, to direct imaging~\cite{2005A&A...438L..29C} and precision atmospheric observations~\cite{2015A&A...576A.134M}, we have now $5.5 \times\, 10^3$ detected exoplanets, and  $\mathcal{O}(10^4)$ candidates awaiting confirmation~\cite{catalog}. Especially with the onset of high performance infra-red astronomy and the JWST~\cite{Brande_2019}, and Roman~\cite{Johnson_2020} space missions low temperature targets, such as exoplanets can be studied across the galaxy, all the way to the Galactic Center (GC). This development opens up a new window to our galaxy, and allows us to study exoplanets in various astrophysical environments. 

Dark Matter (DM) as the dominant mass source in our galaxy is well mapped by stellar kinematics, and despite uncertainties displays a density profile that raises towards the GC~\cite{Benito:2020lgu}. In the search for this ellusive substance a number of studies have considered DM capture in celestial object ranging from the Earth and the Sun~\cite{Batell:2009zp,Pospelov:2007mp,Pospelov:2008jd,Rothstein:2009pm,Chen:2009ab,Schuster:2009au,Schuster:2009fc,Bell_2011,Feng:2015hja,Kouvaris:2010,Feng:2016ijc,Allahverdi:2016fvl,Leane:2017vag,Arina:2017sng,Albert:2018jwh, Albert:2018vcq,Nisa:2019mpb,Niblaeus:2019gjk,Cuoco:2019mlb,Serini:2020yhb,Acevedo:2020gro,Mazziotta:2020foa,Bell:2021pyy}, Jupiter~\cite{Batell:2009zp,Leane:2021tjj,Li:2022wix,French:2022ccb,Ray:2023auh}, Brown Dwarfs~\cite{Leane:2020wob,Leane:2021ihh}, Uranus~\cite{Mitra:2004fh},
White Dwarfs \cite{Garani:2023esk}, Neutron Stars~\cite{Goldman:1989nd,
Gould:1989gw,
Kouvaris:2007ay,
Bertone:2007ae,
deLavallaz:2010wp,
McDermott:2011jp,
Kouvaris:2011fi,
Guver:2012ba,
Bramante:2013hn,
Bell:2013xk,
Bramante:2013nma,
Bertoni:2013bsa,
Kouvaris:2010jy,
McCullough:2010ai,
Perez-Garcia:2014dra,
Bramante:2015cua,
Graham:2015apa,
Cermeno:2016olb,
Graham:2018efk,
Acevedo:2019gre,
Janish:2019nkk,
Krall:2017xij,
Baryakhtar:2017dbj,
Raj:2017wrv,
Bell:2018pkk,
Chen:2018ohx,
Dasgupta:2019juq,
Hamaguchi:2019oev,
Camargo:2019wou,
Bell:2019pyc,
Acevedo:2019agu,
Joglekar:2019vzy,
Joglekar:2020liw,
Bell:2020jou,
Garani:2020wge,
Leane:2021ihh,
Bramante:2021dyx,
Coffey:2022eav},
and other stars~\cite{Freese:2008hb, Taoso:2008kw, Ilie:2020iup, Ilie:2020nzp, Lopes:2021jcy,Ellis:2021ztw,DCandJS}, as well as recently 
Exoplanets~\cite{Leane:2020wob,Bramante:2022pmn} (for a recent review, see~\cite{Bramante:2023djs}). Many of the above studies rely on an energy injection from DM particles that annihilate after being captured by the celestial object. 
This is expected in models where DM has been thermally produced in the early universe. 

Jovian exoplanets have been identified as ideal candidates for DM capture in Ref.~\cite{Leane:2020wob}. The advantage is their relatively large surface gravity, combined with rather low core temperatures, leading to DM with low masses to be retained in the object. Their hydrogen-rich composition makes them kinematically well suited for capturing DM particles around the GeV scale, and especially particles with spin-dependent nuclear interactions. The differential measurement of Jovian planet and Brown Dwarf temperatures close to the GC has been identified as a promising tool to detect heating from annihilating DM ~\cite{Leane:2020wob}. 

In this work we propose a more radical possibility. By investigating the \emph{formation} process of Jovian planets we argue that an additional heat source associated with DM annihilation could halt the formation process at various stages, depending on the DM parameters, such as mass, cross section, and ambient number density. Thus, we would expect to have varying properties of Jovian planets across the galaxy, or their total absence under certain circumstances.

\begin{figure}
\centering
\includegraphics[width=0.95\columnwidth]{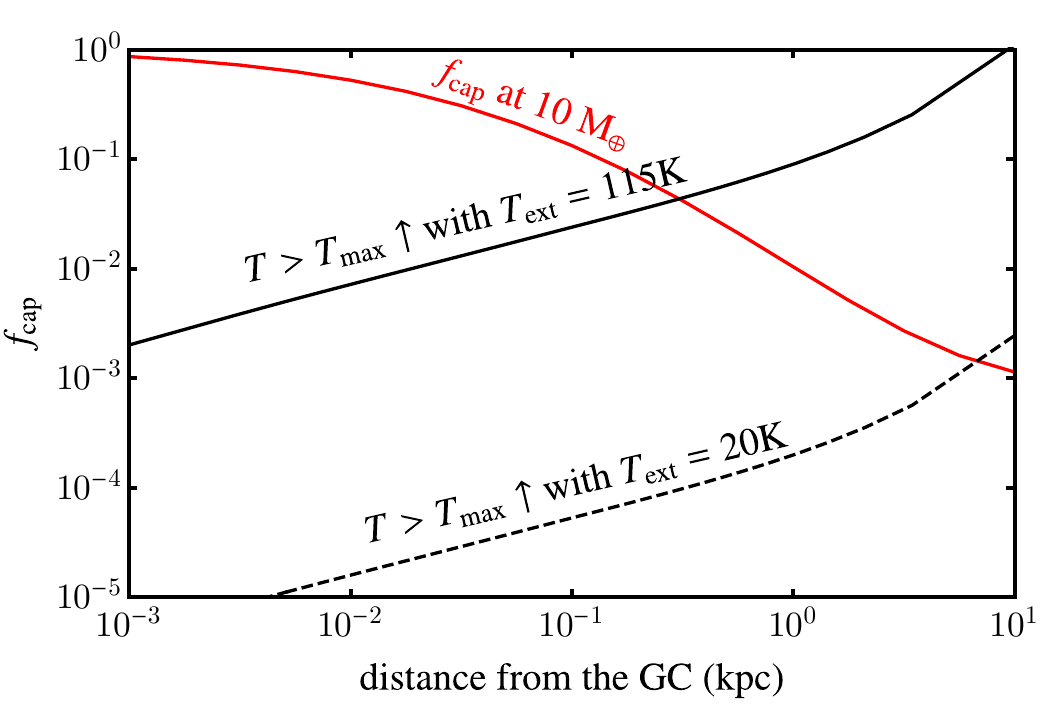}
    \caption{Benchmark for spin-dependent dark matter: $m_{\rm DM} = 10$ GeV, $\sigma_p = 4 \times 10^{-28} \, \rm cm^{2}$. The red line shows the dark matter capture fraction at the time when the planet is $10 \, \rm M_\oplus$ and the black continuous and dashed lines show the fractions needed to halt formation through gas capture. This fraction depends on the temperature of the surrounding gas, and thus on the proximity to the host star. 
    }
    \label{fig:benchplotdist}
\end{figure}

In Fig.~\ref{fig:benchplotdist} we show the capture efficiency as a function of distance from the GC assuming an NFW profile and a Maxwellian velocity dispersion. As shown here, at a given distance from the host star (corresponding to a given external temperature $T_{\rm ext}$), planet formation can be halted in the GC while it does occur some distance away from it. This would imply that towards the GC, Jovian planets appear to occur further closer to the host stars, modulo migration after formation. This observation provides a new and radically different search for a DM heating signal within our galaxy. This distinct phenomenology can be targeted by future observations of exoplanets leveraging the advances in the development of infrared telescopes. 

Finally, the mere existence of Jupiter in our solar system allows us to set new constraints on the parameter space of DM.
We find that future observations of exoplanets near the GC, enabled by the JWST and future infrared telescopes will further test yet unconstrained parameter space of DM candidates across a wide range of masses and interaction cross sections.

\paragraph{Planet formation}
%
We focus on giant, Jovian planets, and the formation process called core-accretion gas-capture~\cite{2005Icar..179..415H}, which proceeds in two stages. First, a solid core made of heavier elements, such as ice and silica forms within a fraction of a Myr. 
These planets form beyond the frost line of a solar system, where icy compounds are cool enough to remain solid, allowing cores to become heavy enough to start to accrete hydrogen and helium.  Planets formed closer to their sun, such as Earth, on the inside the frost line have cores dominated by iron and silicates, which are too rare to lead to large cores that would accrete gas.

Gas capture does not begin until the core accretion exceeds about $10 M_\oplus$ ($ M_\oplus = 6.0 \times 10^{24}$ kg), when the growing core raises the escape velocity above the thermal velocity of the nebular gas. The proto-planet then accumulates a gas envelope at an increasing rate, reflecting its increasing mass. 
The gas can quickly fill up the Bondi sphere of the planet, which subsequently needs to cool via the Kelvin-Helmholtz mechanism in order to accrete further~\cite{pollack1996formation,2022DPS....5410202A}, until eventually the protoplanet reaches a threshold mass at which runaway accretion takes place. Therefore, during the gas accretion period before runaway (which takes place on the scale of a Myr ~\cite{2021Icar..35514087D}) the system remains in a quasi-equilibirum state where gas accretion can only proceed once the excess heat that is released from the gravitational binding energy has been radiated away. 
It is then not surprising that an additional heat source can disrupt this fragile equilibrium and lead to drastically different outcomes. Note in particular that the hydrogen accretion stage is unique, as hydrogen atoms have the lowest gravitational binding energy and are thus most susceptible to evaporation from the forming planet.

\paragraph{Disrupting Jovian Formation}
We assume that DM is produced in a scenario resulting in equal abundances of DM particles and antiparticles.  In the presence of elastic scattering interactions with standard model (SM) particles, DM can scatter in forming planets, lose its kinetic energy, and become gravitationally captured. Once its number density is large enough for particle-anti-particle annihilation to become efficient, it can become an additional substantial heat source in the system. Thus, it may slow or halt Kelvin-Helmholtz contraction, in turn preventing the accretion of gas and the formation of Jovian planets. 
The effect becomes relevant once the energy injection due to DM annihilation is comparable to the power generated by the Kelvin-Helmholtz mechanism.

Contraction is set by the net energy loss, which is modified in the presence of DM annihilation,  
\begin{equation}
\begin{split}\label{eq:enloss}
        \dot{Q} - L_{\rm DM}(R) &= -\frac{d U}{d t} \\ 
        4 \pi R^2 \times \sigma_{\rm SB} T^4 - L_{\rm DM}(R)&= c_1\frac{G M^2}{R^2}\frac{dR}{dt}
\end{split}
\end{equation}
where $\dot{Q} $ is the radiative heat loss, $dU/dt$ the change in gravitational potential energy, $L_{\rm DM} (R)$ is the $R$ dependent heat injection due to DM annihilation, which we call DM luminosity (note that this is not necessarily directly observable as luminosity of the protoplanet), and $1/2 \leq c_1 \leq \infty$
is a factor which depends on the internal mass distribution (for a uniform sphere, $ c_1 = 3/5$). As discussed below DM energy injection leads to halting of the contraction process, such that Eq.(\ref{eq:enloss}) approaches a steady state configuration. We do not consider other subleading effects on the contraction.

Initially, dark matter heat injections will raise the temperature of the protoplanet, increasing its radiative heat loss. However, above a certain temperature the mixture of hydrogen and helium gas acquires enough thermal energy such that particles near its surface reach the escape energy and become unbound. Here, we assume that the system reaches thermal equilibrium fast compared to the time scales of matter accretion. This is justified, as given convection velocities of $v_{\rm conv} \sim 10^{4} \text{cm}/\text{s}$ (see Ref. ~\cite{Freytag:2010kn} Fig. 10) the convection timescale is between 10  and 1000 days, depending on the evolutionary stage.

This process will be in competition with the mass accretion rate, which we take from numerical simulations~\cite{Lissauer:2008hn}. To quantify the particle loss rate we consider the generalisation of the Jeans escape rate $ \Gamma^0_{\rm J}(T)$~\cite{jeans_2009}.
The atmospheric particle loss has a number of uncertainties, based on the distance from the host star, or planetary dynamics, see for example~\cite{yiugit2021martian}. However, for our analytic estimate we choose to focus on the Jeans escape criterion. The big advantage of this simple analytic treatment is that the effect of escape into a dense unbound gas cloud can be included by considering a chemical potential difference, as we demonstrate below. 
\begin{align}
\label{eq:jeansflux}
    & \Gamma_{\rm J} (T)= \Gamma^0_{\rm J} (T) \exp \left(- \frac{\Delta \mu}{ N\, T} \right)  \nonumber \\
    & =  \frac{n v}{2 \sqrt{\pi}} \left(1 + \frac{v_{\rm esc}^2}{v^2} \right) \exp \left(- \frac{E_{\rm esc}}{T} \right) ,
\end{align}
where $n$ is the number density of gas particles, $v_{\rm esc}$ the surface escape velocity,  $v=\sqrt{2T/m}$, and the escape energy that takes into account the difference in chemical potentials
\begin{align}
    E_{\rm esc} = \frac{G_N M m}{r} + \frac{\Delta \mu}{N} = \frac{m\, v_{\rm esc}^2}{2} + \frac{\Delta \mu}{N}\,.
\end{align}
Here, $G_N$ is the Newton constant, $M$ the mass of the forming planet, $m$ the mass of the evaporating particle, and $\Delta \mu/N$ is the difference in chemical potentials per particle. It is intriguing that the chemical potential is relevant in this case, reflecting the special situation of evaporation of particles from the planetary envelope into the surrounding gas of the protoplaneary disk, which has a comparable particle number density and temperature. 

Given that $\mu_1$ is the chemical potential in the envelope and $\mu_2$ that of the protoplaneraty disk, the difference in the chemical potentials $\Delta \mu$ is given by
\begin{align}
    \mu_1 -\mu_2 = U_1-U_2 + P_1 V_1 - P_2 V_2  + T_2 S_2 - T_1 S_1 
\end{align}
where we assumed that the gravitational energies of the particles are of the same order. We further assume that the particles behave as ideal gas, and thus the internal energy is $U_i = C_V N_i T_i$, $P_i V_i = N_i T_i$, and the entropy is given by the Sackur-Tetrode formula $S_i = N_i \kappa_i$, with
\begin{align}
 \kappa_i =   \left( \frac{5}{2} + \log \left[  \frac{V_i}{N_i} \left( \frac{4 \pi m U_i}{3 N_i} \right)^{3/2} \right] \right)\,.
\end{align}
Furthermore, we assume that the gases are monoatomic, and hence the heat capacity $C_V \approx 3/2$. 

We can therefore express the chemical potential difference per envelope particle as 
\begin{align}
&\frac{\Delta \mu }{N_1} = C_V \left( T_1 - \frac{N_2}{N_1} T_2\right) + T_1 - \frac{N_2}{N_1} T_2 + \bar{\kappa} \left( T_2 \frac{N_2}{N_1} -  T_1  \right)  \nonumber \\
&  \approx \left( 1 + C_V - \bar{\kappa}\right) \left(T_1 - T_2\right) = \frac{3}{2} \log \left[    \frac{4 \pi m \bar{u}}{3  \bar{n}^{2/3}} \right] \left( T_2 - T_1\right) \,.
\end{align}
Where we have used the assumption that $N_1 \approx N_2$, and as the logarithm is a slowly varying function, the argument can be replaced by the average value of the internal energy per particle $\bar{u} = \bar{U}/\bar{N} \approx C_V \bar{T}$. Thus, we observe that in the case that the surrounding gas is hotter than the envelope an additional suppression slows particle evaporation down. 

Now we compare the escape rate with the rate of growth of the planet in the absence of dark matter heat $\dot{M} $, which we take from simulation results presented in~\cite{Lissauer:2008hn}. Equating those rates
\begin{align}
A_{\rm planet}\, \Gamma_{\rm J} (T) |_{T=T_{\rm max}} = \mathcal{O}(1) \, \dot{M}    
\end{align}
where $A_{\rm planet}$ is the surface from which evaporation takes place, we find that the result is insensitive to the value of the numerical $\mathcal{O}(1)$ factor in the equation, since it can not compete with the exponential in the escape rate (see also \cite{d2021growth}, which is largely consistent but presented slower accretion, implying smaller DM injections are needed to match it). 

For an ambient temperature of $115$ K of the surrounding gas~\cite{Lissauer:2008hn}, this treatment gives us hydrogen escape temperatures of $ T_{\rm max} \sim 85 $ K  for the envelope. For the core the assumption above has to be modified given that the particle number density in the core is much higher than in the surrounding gas. We get  $ T_{\rm max} \sim 10^3 $ K for the core, a result that is not significantly affected by the ambient gas properties for the systems we study.

Thus, the maximum heat loss happens just below $T_{\rm max}$ and we estimate that DM heat injection halts further gas capture if
\begin{equation}
\label{eq:haltcondition}
\begin{split}
    L_{\rm DM} &\geq 4 \pi R^2 \sigma T_{\rm max}^4 \\
    &\geq 4 \times 10^{-4} 
    \times \left(\frac{R}{10^3 R_p} \right)^{2} \, 
    \left(\frac{T_{\rm max}}{ 80 \rm K} \right)^{4}
    \rm \, L_\odot   
\end{split}
\end{equation}
where $R_p = 7 \times 10^5$ km is the radius of Jupiter (the protoplenatary envelope has radius $\sim 10^3 R_p $), and where $L_\odot = 2 \times 10^{36} \, \rm GeV \, s^{-1}$ is the luminosity of the Sun.
To investigate whether Eq.~\eqref{eq:haltcondition} is satisfied in the evolution of the planet, we use the simulation results presented in \cite{d2021growth}.
We will now examine how the DM luminosity depends on the properties of the capturing object, and the DM particle model.

\paragraph{Dark matter energy injection}
To obtain the DM luminosity, we assume DM annihilation equilibrium -- motivated a posteriori below -- i.e. that each incoming DM particle that is captured contributes its rest mass immediately to the energy budget. We will discuss why this treatment is justified in the following subsection.  We thus can write
\begin{equation} \label{eq:Lanneq}
    L_{\rm DM} = m_{\rm DM} C_{\rm cap} =   m_{\rm DM} f_{\rm cap} \Phi \pi R^2
 \end{equation}
where $C_{\rm cap}$ is the capture rate, and $f_{\rm cap}$ the captured fraction of the DM flux  $\Phi$  passing through the object, given by
\begin{equation}
    \Phi =   \, v_{\rm DM} \sqrt{\frac{8}{3 \pi}} \frac{\rho_{\rm DM}}{m_{\rm DM}} \left(1 + \frac{3}{2} \frac{v_{\rm esc}^2}{v_{\rm DM}^2} \right) \,. 
\end{equation}
Then, we have
\begin{equation}
    \begin{split}
        L_{\rm DM}
        \sim& 7\times 10^{-4} f_{\rm cap} 
        \left(\frac{R}{10^3 R_p} \right)^2 
        \\ &\times \left( \frac{v_{\rm DM}}{270 \rm km \, s^{-1}}\right) \left(\frac{\rho_{\rm DM}}{0.42 \, \rm GeV \, cm^{-3}} \right)
        \,L_\odot
    \end{split}
\end{equation}
which only depends on the dark matter mass through $f_{\rm cap}$. It is seen that this may indeed exceed \eqref{eq:haltcondition}. 

\subparagraph{Dark matter capture}
Given a certain DM mass and velocity, the DM particle needs a certain number of scatterings to drop below the escape velocity of a given object. This will in general also depend on the efficiency of the energy loss, determined by the mass ratio of the DM particle and the target nucleus -- DM is not sufficiently slowed down when this mass ratio is too large.  Furthermore, in the case of DM particles with masses similar to or below the target nucleus mass, DM particles might be reflected off the object before capture can occur, as discussed in detail in Ref.~\cite{RKLandJS,asteria}. To compute the fraction of the DM particles that are captured in a given object we use the \verb|Asteria|~\cite{asteria, RKLandJS} package, which takes all the above mentioned phenomena into account.

In this work we focus on DM interaction with nuclei and distinguish two broad DM model classes. The first is a model where the DM particle scatters with SM particles independently of their spin. Given an input DM nucleon cross section $\sigma_{\chi N}$, this leads to the spin-independent DM-nucleus cross section with an atomic number $A$ nucleus 
\begin{align}
\sigma_{\rm SI} = \sigma_{\chi N}^{\rm SI} A^2 \, \left( \frac{\mu(m_\chi,m_A )}{\mu(m_\chi,m_N)} \right)^2 \, ,
\end{align} 
where $\mu(m_\chi, m_{\rm SM})$ is the reduced mass of a DM and a SM particle of mass $m_{\rm SM}$. Note that it has been pointed out that at cross sections $\sigma_{\chi N} > 10^{-26} \text{cm}^2$ the scattering is not described by interactions of point like particles, and the coherent scattering picture breaks down \cite{Digman:2019wdm,Cappiello:2020lbk}. Rather, at cross sections larger than this threshold the interaction can be assumed as a total geometric scattering cross section with the entire nucleus without an implied scaling with atomic number. 

The second case of interactions we consider is a spin-dependent scattering cross section
\begin{align}
    \sigma_{SD} = \sigma_{\chi N}^{\rm SD} \, \frac{4 \left(J_A + 1\right)}{3 J_A} \large[ a_p \langle S_p \rangle + a_n \langle S_n \rangle  \large]^2\, \left( \frac{\mu(m_\chi,m_A )}{\mu(m_\chi,m_N)} \right)^2 \, , 
\end{align}
where $J_A$ is the nuclear spin of the target, $\langle S_p \rangle$ the average proton and $\langle S_n \rangle$ the average neutron spin of the nucleus. The $a_p$, and $a_n$ are the DM model parameters indicating the coupling strength to protons and neutrons respectively. For example in the case of $a_p = 1$, and $a_n = 0$, the hydrogen scattering cross section is equal to the DM scattering cross section on protons $\sigma_{SD} = \sigma_p$, which we will use as a benchmark model.
 
To compute the capture of DM in the protoplanet we introduce a simplifying effective framework and use the \verb|Asteria|~\cite{asteria, RKLandJS} package to compute capture in two regimes depending on where in the object we expect the DM to be captured. 
In the first regime the DM capture takes place on the scale of the entire object and the average chemical composition of the protoplanet is taken as an input for \verb|Asteria|, while in the other regime the capture takes place in the envelope. In the second case only the chemical composition of the envelope is used, but the total mass of the protoplanet determines the escape velocity. To distinguish the regimes we define the effective capture length  $\lambda_{\rm cap}$ and compare it to the size of the protoplanet, which is dominated by the size of the envelope. 
\begin{align}
    & \lambda_{\rm cap} = \lambda_{\rm mfp} \, N_{\rm cap} \text{ with } \lambda_{\rm mfp}  = \left( \sigma_{\rm SI/SD}\,  n_{\rm SM} \right)^{-1} \, ,\\ \nonumber
   &  N_{\rm cap} = \frac{\log \left( v_{\text{esc}}^2 / (v_\chi^2 + v_{\text{esc}}^2) \right)}{\log \left( 1 - 2 \mu (1-\mu)^{-2}\right)} \,,
\end{align}
where $N_{\rm cap}$ is the number of scatterings required to capture a DM particle~\cite{Mack:2007xj},  $v_{\text{esc}}$ the protoplanets' escape velocity, $v_{\chi}$ the DM impact velocity, and $\mu = m_{\rm DM}/m_{\rm SM}$ is the mass ratio between the DM and the average SM particle in the protoplanet. Note that this estimate provides an upper bound on the capture length of DM as a straight-line trajectory is assumed.

To determine the chemical composition we model the protoplanet in a simplified manner assuming that it consists of an icy core and an outer gas envelope, both are assumed to have a homogeneous average density. We use the mass and radius from Ref.~\cite{Lissauer:2008hn} as time dependent input quantities.  
The outer envelope has a large radius and a low density. 
We model the envelope to consist of 75\% hydrogen and 25\% helium. 
The core is assumed to have a $58/42$  ice ($\rm H_2 O$) to silica ($\rm Si O_2$) mass ratio~\cite{2021Icar..35514087D}, which makes Oxygen the most abundant element. This implies that in the regime where the core contributes to the capture, optimal capture happens for slightly heavier DM candidates, due to more efficient energy transfer between the SM and DM particles of similar mass.

For the spin-dependent calculation, we need to estimate the mass fraction which is relevant for spin-dependent scattering. Thus we take the isotope fractions and properties of the elements in Tab.~\ref{table:isotopes}.
\begin{table}
\centering
\caption{Isotope fractions, nuclear spins, average proton, and neutron spins of the dominant isotopes considered. Note that the quoted abundances are relative isotope fractions of the considered isotopes.}
\label{table:isotopes} 
\begin{tabular}{|c|c|c|c|}
\hline
\textbf{Isotopes} & \textbf{\ce{^{17}O}} & \textbf{\ce{^{29}Si}} & \textbf{\ce{^{1}H}} \\
\hline\hline 
Abundance [$\%$] & $\sim 0.4$ & $\sim 4.7$ & $\sim 100$ \\
\hline
\textbf{J} & 5/2 & 1/2 & 1/2 \\
\hline
$\langle S_p \rangle$ & -0.036 & 0.054 & 0.5 \\
\hline
$\langle S_n \rangle$ & 0.508 & 0.204 & 0 \\
\hline
\end{tabular}
\end{table}

Lastly, we note that light DM candidates can be reflected, such that the capture efficiency remains below one even at large cross sections~\cite{RKLandJS}. On the other hand, in models with a long range force mediator, the escape velocity in the medium can be increased, which in turn suppressed the reflection probability~\cite{Acevedo:2023owd}. We discuss the implications for the cross section sensitivity estimate in the following section.

\subparagraph{Dark matter annihilation}
Neglecting DM evaporation from the celestial body, the Boltzmann equation governing the accumulation of DM is given by
\begin{align}
\frac{dN}{dt} = C_{\rm cap} - C_{\rm ann} N_\chi^2 \, ,
\end{align}
where $C_{\rm cap}$ is the total capture rate of DM particles per unit time in the celestial object, and $C_{\rm ann} =  \langle \sigma_{\rm ann} v \rangle / V_{\rm eff}$ the DM annihilation rate per effective volume, which is given by $1/V_{\rm eff} = \int_V n_\chi^2/\int_V n_\chi$. After a characteristic time $\tau$ the system reaches annihilation equilibrium where $dN_{\chi}/dt \rightarrow 0$, which implies that  $N_\chi \rightarrow \sqrt{C_{\rm cap} / C_{\rm ann}}$, leading to $ \tau = 1/\sqrt{ C_{ \rm cap} C_{\rm ann}}$. 

For our forming protoplanet the $V_{\rm eff} \approx V_{\rm core}$, which typically has a $R \sim 3\, R_\oplus$, we thus have 
\begin{align}
\tau \approx \sqrt{\frac{ m_\chi \, 10^{-30} \text{ cm}^3/\text{s} }{ \langle \sigma_{\rm ann} v\rangle \text{ GeV} }} \text{ Myr}\,,
\end{align}
which for DM masses around a GeV, and typical thermal annihilation cross sections is below the Myr time-scale. We therefore assume that the protoplanet reaches annihilation equilibrium, and the total rate of captured DM particles is directly converted to the DM luminosity, as discussed above.

\begin{figure}[t!]
\centering
\includegraphics[width=0.95\columnwidth]{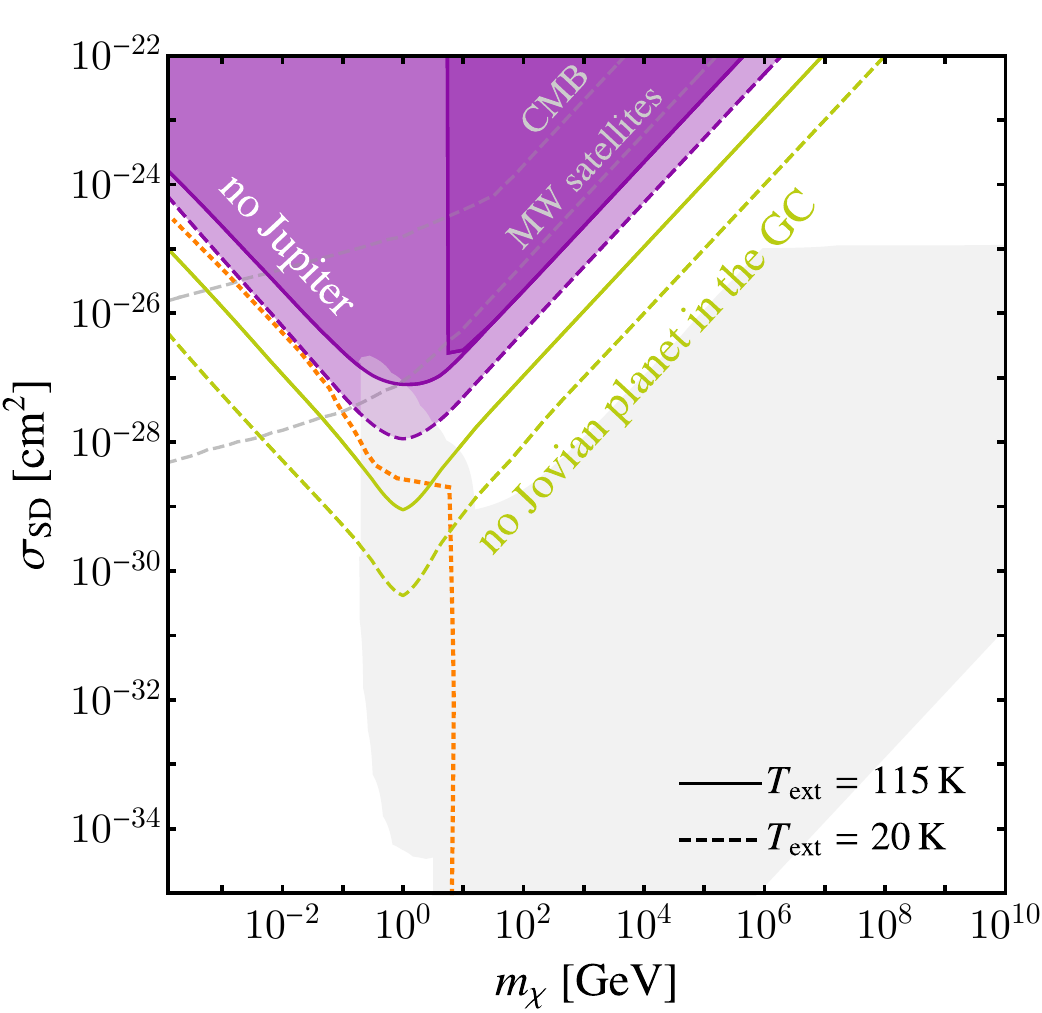}
    \caption{Constraints on spin-dependent parameter space. The filled purple contours demonstrate the cross sections for which the formation of Jupiter stalls at $ 10 \rm \, M_\oplus$, before gas is captured (see text for motivation of this benchmark).
    The green contours demonstrate the cross sections for which the formation of Jovian planets near the GC stalls at this same mass.
    Continuous contours assume that the environment in which the planet forms is $115 \rm K$; dashed contours assume it is $20 \rm K$.
    Reflection of light DM is considered for all but the outer purple $T=115 \rm K$ contour. As we discussed above reflection is a model dependent phenomenon, and can be strongly suppressed in models with a long range force mediator.  
    Finally, in the region below the orange dotted line DM evaporates from the forming planet as discussed in appendix \ref{sec:evap}, given short-range interactions only. 
    We also show constraints from direct detection (filled gray contours) and from the CMB \cite{Boddy:2018kfv} and MW-satelites \cite{Nadler:2019zrb} (dashed lines).  
}
    \label{fig:constraintsSD}
\end{figure}
\begin{figure}[t!]
\centering
\includegraphics[width=0.95\columnwidth]{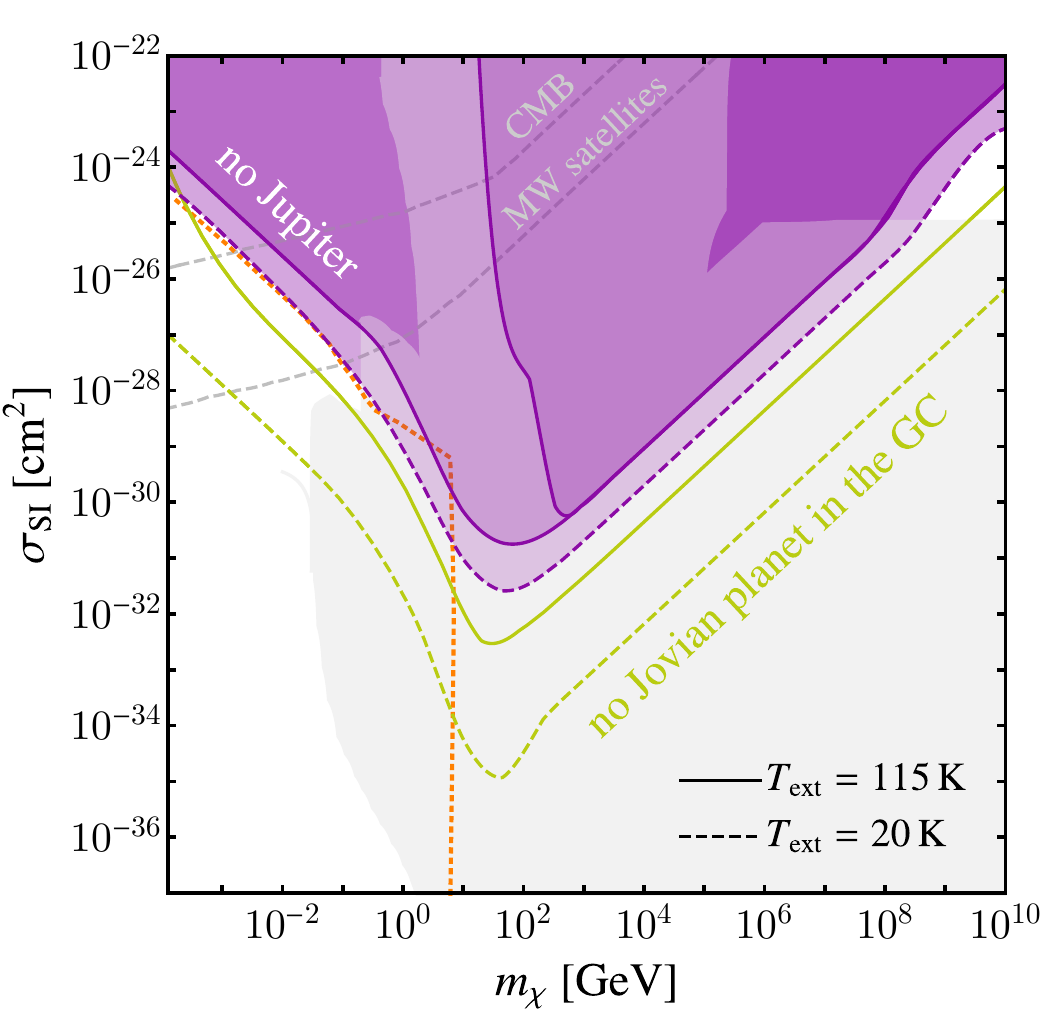}
\caption{Constraints on spin-independent DM interactions from the existence of Jupiter, and expected sensitivity from the non-existence of Jovians in the GC, as well as direct detection and cosmology dependent constraints -- as in Fig.\ref{fig:constraintsSD}.
}
\label{fig:constraintsSI}
\end{figure}

\paragraph{Implications}
\label{sec:implications}
We study the DM fraction needed to halt the formation process for Jupiter in the solar system, using $ \rho_{\rm DM} = 0.42 \, \rm GeV \, cm^{-3}$ and $ v_{\rm DM} = 270 \rm km \, s^{-1}$, and find that $f_{\rm cap}$ remains approximately constant for the period of formation that we are interested in.

In Figs.~\ref{fig:constraintsSD} and \ref{fig:constraintsSI} we show constraints on the 
spin-dependent and spin-independent scattering cross sections respectively from the fact that the formation of Jupiter in our solar system has not stopped at a mass of $10\rm M_\oplus$ as purple contours. 
As the temperature increase needed to satisfy \eqref{eq:jeansflux} does grow as a function of time, 
we expect that formation of Jovian planets in the GC halts at $10 \rm M_\oplus$ in the parameter range indicated by the contours, and proceeds as normal below these contours.
We also project constraints based on the observation of Jovian planets in the GC (for which we have assumed $\rho_{\rm DM} = 10^3 \, \rm GeV \, cm^{-3}$ and $v_{\rm DM} = 10 \rm \, km \, s^{-1}$) in chartreuse (note that the premature ending is already in tension with inferences of Jovian planets in the GC from lensing events \cite{Sahu:2006ex,MOA:2009dkq,koshimoto2021no}).
We show these constraints and projections for two different benchmarks for the protostellar disk temperature; the more conservative $115$ K and $20$ K, for planets forming further away from the star. We note that this is not the only parameter degeneracy; most notably, the unavailability of nearby gas and dust can also cause a protoplanet to stop accreting, and the migration of a planet away from the star can likewise change its accretion rate. Therefore, while one can interpret the observation of a large gas giant as a constraint on DM, the non-observation cannot directly be interpreted as positive evidence. 

For comparison we show current direct detection constraints as gray contours. Here we show the maximal reach in the sensitivity to the direct nuclear recoil process, which are least model dependent.  The reach is dominated by the CRESST results at lower masses for spin-independent~\cite{CRESST:2022lqw}, and spin-dependent~\cite{CRESST:2022dtl} cross sections, while at higher masses the LUX~\cite{LUX:2017ree} results dominate for spin-dependent interactions, while the XENON experiment dominates the spin-independent sensitivity~\cite{XENONCollaboration:2023orw}. For large spin-independent cross sections, we also show the XQC constraint \cite{Erickcek:2007jv,Mahdawi:2018euy}. 
For small masses and spin-independent cross section, the SuperCDMS Migdal constraints are shown \cite{SuperCDMS:2023sql}. We do note that these results have large theoretical uncertainties~\cite{Cox:2022ekg}.
We do not show the constraints resulting from the observation of the heat flow from the Earth, as those studies neglected the DM evaporation, and thus the contours are not accurate at DM masses below $\mathcal{O}(10)$ GeV, however at larger DM masses the limits coincide with our findings~\cite{Mack:2007xj, Bramante:2019fhi}

The sensitivity ceiling of direct detection experiments at large cross sections for light DM from Ref.~\cite{Cappiello:2023hza} and for heavy DM from Refs.~\cite{Kavanagh:2017cru, Cappiello:2020lbk} is shown.  Since at the large cross sections considered the scaling with atomic mass number is uncertain~\cite{Digman:2019wdm}, we do not rescale the results for spin-dependent interactions and a more detailed analysis to obtain a precise ceiling values is needed. 
Note that in contrast the sensitivity of our constraints and suggested search is only affected at cross sections above $10^{-22} \text{cm}^2$, where large drift times would require a more detailed analysis of the DM distribution~\cite{Acevedo:2020gro,Leane:2022hkk} and annihilation outside the core.

We also show complementary cosmology-dependent constraints from the CMB \cite{Boddy:2018kfv}. DM which scatters with a velocity-independent cross-section can also be constrained from the lack of MW satellites produced in such models \cite{Nadler:2019zrb}. However, as the DM scattering rates could be different in the early universe~\cite{Elor:2021swj}, those bounds are not as universal as the shown late-time sensitivity.

Lastly, as a result of DM heat injection, the proto-planet will respond by radiating away more energy. Even if the heating effect is not significant enough to halt the formation of the gas envelope, this effect leads to an increase in luminosity in high DM-density environments. The luminosity of protoplanet formation peaks in the near-infrared and can be determined by future observations. 
This allows a direct comparison of the observed peak luminosity, realised at about $2$ Myr, and that predicted by theoretical models.

\paragraph{Discussion}
In this paper, we have studied the inhibition of the formation of Jupiter in our solar system and Jovian planets in the Milky Way galaxy by the capture and annihilation of DM. We have shown that the heat generated by DM annihilation can increase the temperature of the protoplanet above the temperature at which the gas mixture evaporates, and determined the corresponding constraints on the parameter space of spin-(in)dependent DM. 
These constraints are competitive for relatively strongly interacting DM candidates, above the ``ceiling'' of direct detection constraints. We have shown that these constraints depend on the ambient temperature at formation, with stronger effects for planets forming further away from the star.

We have also estimated the projected sensitivities from the existence of Jovian planets in the GC, where the DM density is higher and the typical velocity dispersion lower. Future expolanet observations by JWST and Roman may be used to test these scenarios. Moreover, as we explain in section~\ref{sec:implications}, for a particular DM (sub-component) mass and cross section the formation of Jovian planets is affected as a function of the distance from the GC, enabling a differential detection of the heating signal.

In this work, we have estimated the sensitivity of protoplanets to DM injections using a semi-analytical calculation, with input from simulations in the absence of these injections. Future work should focus on the embedding of DM heat in planetary formation simulations, such that more realistic projections can be obtained.
Here we have focused on  the core-accretion gas-capture theory of giant planet formation. This is the simplest scenario for the formation of Jovian planets. Alternative scenarios include dynamical processes and migration within the stellar system. Since we focus on the final stages of formation, we expect such alternative scenarios to be qualitatively similar.

\section*{Acknowledgements}
We thank Chris Cappiello, Malcolm Fairbairn, Scott Gaudi, Rebecca Leane, Dave McKeen, and Nirmal Raj for useful conversations and feedback on the manuscript. DC is supported by the STFC under Grant No. ST/T001011/1.

\appendix
\section{Dark matter evaporation}
\label{sec:evap}
The condition \eqref{eq:haltcondition} relies on the fact that hydrogen evaporates for temperatures above $ T_{\rm max}$. Therefore, one would naively say that DM candidates with masses below the mass of hydrogen would evaporate at these temperatures too. However, unlike hydrogen, as we demonstrate below the DM particles are concentrated in the core of the objects. Thus, in the short mean free path regime the lower number density of DM particles in the envelope implies a lower flux of escaping particles \eqref{eq:jeansflux}. Here we estimate the evaporation rate of DM particles of a given mass, and compare it to the capture rate.

The DM number density within the protoplanet is obtained by integrating the radial equation Eq. (11) form Ref.~\cite{Leane:2022hkk}, under the assumption that the envelope has constant density $\rho_E$ (at $t=\rm Myr$, $\sim 3\times 10^{-8} \rm g \, cm^{-3}$) and the core has a constant density $\rho_{\rm core}$ (at $t= \rm Myr$, $\sim 2 \rm g \, cm^{-3}$). We distinguish two regimes, the isothermal regime, where the DM mean free path $\lambda$ in the object is of the order of the object, and DM thermalises with the core at a constant temperature. The second regime where local thermal equilibrium (LTE) is applicable when the DM mean free path is smaller than the size of the object. Thus we have the total DM profile:
\begin{align}
    n_{\chi} =  
      \begin{cases}
      g(r) /N_0  \, \exp \left(  - F(r)\right) \quad \text{if } \lambda < R_E \\ 
    1/N_0 \exp(-m_\chi  \Phi(r)/T(r))  \quad \text{if } \lambda > R_E
        \end{cases}
    \, , 
\end{align}
where the radius is normalised to the envelope size $r = R/R_E$, $\Phi(r)$ is the gravitational potential, $T(r)$ the temperature profile, the normalisation factor $N_0$ ensured that we have $N_\chi =  4 \pi \int_0^{R_E} r^2 n_\chi(r)$,  and the LTE functions are given by 
\begin{align}
& \frac{F(r) }{   \frac{4}{3} \pi  G_N  m_\chi  R_E^2   }  =      \frac{r^2 \rho_{\rm core} }{ T_{\rm core}} \theta \left(\frac{ R_{\rm core} }{ R_E}-r\right) +   \theta \left(r- \frac{ R_{\rm core} }{ R_E}  \right)      \nonumber \\ 
&  \times   \frac{   \left(1/2\,  r^2 \rho_E  + \left( \frac{ R_{\rm core} }{ R_E } \right) ^3 ( \rho_{\rm core} -  \rho_E ) (  \frac{R_E}{ R_{\rm core}} -  \frac{1}{r}) \right)}{ T_E} \, ,    
\end{align}

\begin{align}
& g(r) = \theta \left(r-\frac{R_{\rm core}}{R_E}\right) \left(\frac{T_E}{T_{\rm core}}\right){}^{    \frac{1}{2}  \left(\frac{m_{\chi }}{m_{\text{p}}}+1\right)^{-3/2}-1}   \nonumber  \\ 
&  +\theta \left(\frac{R_{\rm core}}{R_E}-r\right) \,,
\end{align}
where $ \theta$ indicates the Heaviside function.
For the regime where the mean free path of DM becomes comparable to SM particles the above equation provides us with the DM density within the last scattering surface, to which we apply the Jeans' escape criterion, demanding that the total surface evaporation rate is below the capture rate. While for the isothermal, and transition regime we use the formalism of Ref.~\cite{1987ApJ...321..560G, 1990ApJ...356..302G} to determine the evaporation mass comparing the evaporation rate and the time-scale needed to reach annihilation equilibrium assuming a core temperature of $3000$ K. This leads to the evaporation contour, shown by the dotted lines in Figs.~\ref{fig:constraintsSD},~\ref{fig:constraintsSI}. Note that this line is only marking to lowest mass of the DM retained in the object if DM interacts through contact interactions only. In the presence of attractive long range forces, the evaporation contours are significantly moved to lower DM masses, as discussed in Ref.~\cite{Acevedo:2023owd}.

\bibliography{refs}

\end{document}

%% file: universalnewcommands.tex
\newcommand{\beq}{\begin{equation}}
\newcommand{\eeq}{\end{equation}}
\newcommand{\bea}{\begin{eqnarray}}
\newcommand{\eea}{\end{eqnarray}}

\newcommand{\refjnl}[1]{{\rm#1}}

\def\apj{\refjnl{ApJ}}                 
\def\aap{\refjnl{A\&A}}                

\def\icarus{\refjnl{Icarus}}           

\def\nat{\refjnl{Nature}}              



%% file: mainV3.bbl
\begin{thebibliography}{119}%
\makeatletter
\providecommand \@ifxundefined [1]{%
 \@ifx{#1\undefined}
}%
\providecommand \@ifnum [1]{%
 \ifnum #1\expandafter \@firstoftwo
 \else \expandafter \@secondoftwo
 \fi
}%
\providecommand \@ifx [1]{%
 \ifx #1\expandafter \@firstoftwo
 \else \expandafter \@secondoftwo
 \fi
}%
\providecommand \natexlab [1]{#1}%
\providecommand \enquote  [1]{``#1''}%
\providecommand \bibnamefont  [1]{#1}%
\providecommand \bibfnamefont [1]{#1}%
\providecommand \citenamefont [1]{#1}%
\providecommand \href@noop [0]{\@secondoftwo}%
\providecommand \href [0]{\begingroup \@sanitize@url \@href}%
\providecommand \@href[1]{\@@startlink{#1}\@@href}%
\providecommand \@@href[1]{\endgroup#1\@@endlink}%
\providecommand \@sanitize@url [0]{\catcode `\\12\catcode `\$12\catcode
  `\&12\catcode `\#12\catcode `\^12\catcode `\_12\catcode `\%12\relax}%
\providecommand \@@startlink[1]{}%
\providecommand \@@endlink[0]{}%
\providecommand \url  [0]{\begingroup\@sanitize@url \@url }%
\providecommand \@url [1]{\endgroup\@href {#1}{\urlprefix }}%
\providecommand \urlprefix  [0]{URL }%
\providecommand \Eprint [0]{\href }%
\providecommand \doibase [0]{http://dx.doi.org/}%
\providecommand \selectlanguage [0]{\@gobble}%
\providecommand \bibinfo  [0]{\@secondoftwo}%
\providecommand \bibfield  [0]{\@secondoftwo}%
\providecommand \translation [1]{[#1]}%
\providecommand \BibitemOpen [0]{}%
\providecommand \bibitemStop [0]{}%
\providecommand \bibitemNoStop [0]{.\EOS\space}%
\providecommand \EOS [0]{\spacefactor3000\relax}%
\providecommand \BibitemShut  [1]{\csname bibitem#1\endcsname}%
\let\auto@bib@innerbib\@empty
\bibitem [{\citenamefont {{Mayor}}\ and\ \citenamefont
  {{Queloz}}(1995)}]{1995Natur.378..355M}%
  \BibitemOpen
  \bibfield  {author} {\bibinfo {author} {\bibfnamefont {M.}~\bibnamefont
  {{Mayor}}}\ and\ \bibinfo {author} {\bibfnamefont {D.}~\bibnamefont
  {{Queloz}}},\ }\href {\doibase 10.1038/378355a0} {\bibfield  {journal}
  {\bibinfo  {journal} {\nat}\ }\textbf {\bibinfo {volume} {378}},\ \bibinfo
  {pages} {355} (\bibinfo {year} {1995})}\BibitemShut {NoStop}%
\bibitem [{\citenamefont {{Chauvin}}\ \emph {et~al.}(2005)\citenamefont
  {{Chauvin}}, \citenamefont {{Lagrange}}, \citenamefont {{Zuckerman}},
  \citenamefont {{Dumas}}, \citenamefont {{Mouillet}}, \citenamefont {{Song}},
  \citenamefont {{Beuzit}}, \citenamefont {{Lowrance}},\ and\ \citenamefont
  {{Bessell}}}]{2005A&A...438L..29C}%
  \BibitemOpen
  \bibfield  {author} {\bibinfo {author} {\bibfnamefont {G.}~\bibnamefont
  {{Chauvin}}}, \bibinfo {author} {\bibfnamefont {A.~M.}\ \bibnamefont
  {{Lagrange}}}, \bibinfo {author} {\bibfnamefont {B.}~\bibnamefont
  {{Zuckerman}}}, \bibinfo {author} {\bibfnamefont {C.}~\bibnamefont
  {{Dumas}}}, \bibinfo {author} {\bibfnamefont {D.}~\bibnamefont {{Mouillet}}},
  \bibinfo {author} {\bibfnamefont {I.}~\bibnamefont {{Song}}}, \bibinfo
  {author} {\bibfnamefont {J.~L.}\ \bibnamefont {{Beuzit}}}, \bibinfo {author}
  {\bibfnamefont {P.}~\bibnamefont {{Lowrance}}}, \ and\ \bibinfo {author}
  {\bibfnamefont {M.~S.}\ \bibnamefont {{Bessell}}},\ }\href {\doibase
  10.1051/0004-6361:200500111} {\bibfield  {journal} {\bibinfo  {journal}
  {\aap}\ }\textbf {\bibinfo {volume} {438}},\ \bibinfo {pages} {L29} (\bibinfo
  {year} {2005})},\ \Eprint {http://arxiv.org/abs/astro-ph/0504658}
  {arXiv:astro-ph/0504658 [astro-ph]} \BibitemShut {NoStop}%
\bibitem [{\citenamefont {{Martins}}\ \emph {et~al.}(2015)\citenamefont
  {{Martins}}, \citenamefont {{Santos}}, \citenamefont {{Figueira}},
  \citenamefont {{Faria}}, \citenamefont {{Montalto}}, \citenamefont
  {{Boisse}}, \citenamefont {{Ehrenreich}}, \citenamefont {{Lovis}},
  \citenamefont {{Mayor}}, \citenamefont {{Melo}}, \citenamefont {{Pepe}},
  \citenamefont {{Sousa}}, \citenamefont {{Udry}},\ and\ \citenamefont
  {{Cunha}}}]{2015A&A...576A.134M}%
  \BibitemOpen
  \bibfield  {author} {\bibinfo {author} {\bibfnamefont {J.~H.~C.}\
  \bibnamefont {{Martins}}}, \bibinfo {author} {\bibfnamefont {N.~C.}\
  \bibnamefont {{Santos}}}, \bibinfo {author} {\bibfnamefont {P.}~\bibnamefont
  {{Figueira}}}, \bibinfo {author} {\bibfnamefont {J.~P.}\ \bibnamefont
  {{Faria}}}, \bibinfo {author} {\bibfnamefont {M.}~\bibnamefont {{Montalto}}},
  \bibinfo {author} {\bibfnamefont {I.}~\bibnamefont {{Boisse}}}, \bibinfo
  {author} {\bibfnamefont {D.}~\bibnamefont {{Ehrenreich}}}, \bibinfo {author}
  {\bibfnamefont {C.}~\bibnamefont {{Lovis}}}, \bibinfo {author} {\bibfnamefont
  {M.}~\bibnamefont {{Mayor}}}, \bibinfo {author} {\bibfnamefont
  {C.}~\bibnamefont {{Melo}}}, \bibinfo {author} {\bibfnamefont
  {F.}~\bibnamefont {{Pepe}}}, \bibinfo {author} {\bibfnamefont {S.~G.}\
  \bibnamefont {{Sousa}}}, \bibinfo {author} {\bibfnamefont {S.}~\bibnamefont
  {{Udry}}}, \ and\ \bibinfo {author} {\bibfnamefont {D.}~\bibnamefont
  {{Cunha}}},\ }\href {\doibase 10.1051/0004-6361/201425298} {\bibfield
  {journal} {\bibinfo  {journal} {\aap}\ }\textbf {\bibinfo {volume} {576}},\
  \bibinfo {eid} {A134} (\bibinfo {year} {2015})},\ \Eprint
  {http://arxiv.org/abs/1504.05962} {arXiv:1504.05962 [astro-ph.EP]}
  \BibitemShut {NoStop}%
\bibitem [{\citenamefont {{NASA}}(2020)}]{catalog}%
  \BibitemOpen
  \bibfield  {author} {\bibinfo {author} {\bibnamefont {{NASA}}},\ }\href@noop
  {} {\enquote {\bibinfo {title} {Nasa exoplanet catalog},}\ }\bibinfo
  {howpublished} {\url{https://science.nasa.gov/exoplanets/}} (\bibinfo {year}
  {2020})\BibitemShut {NoStop}%
\bibitem [{\citenamefont {Brande}\ \emph {et~al.}(2019)\citenamefont {Brande},
  \citenamefont {Barclay}, \citenamefont {Schlieder}, \citenamefont {Lopez},\
  and\ \citenamefont {Quintana}}]{Brande_2019}%
  \BibitemOpen
  \bibfield  {author} {\bibinfo {author} {\bibfnamefont {J.}~\bibnamefont
  {Brande}}, \bibinfo {author} {\bibfnamefont {T.}~\bibnamefont {Barclay}},
  \bibinfo {author} {\bibfnamefont {J.~E.}\ \bibnamefont {Schlieder}}, \bibinfo
  {author} {\bibfnamefont {E.~D.}\ \bibnamefont {Lopez}}, \ and\ \bibinfo
  {author} {\bibfnamefont {E.~V.}\ \bibnamefont {Quintana}},\ }\href {\doibase
  10.3847/1538-3881/ab5444} {\bibfield  {journal} {\bibinfo  {journal} {The
  Astronomical Journal}\ }\textbf {\bibinfo {volume} {159}},\ \bibinfo {pages}
  {18} (\bibinfo {year} {2019})}\BibitemShut {NoStop}%
\bibitem [{\citenamefont {Johnson}\ \emph {et~al.}(2020)\citenamefont
  {Johnson}, \citenamefont {Penny}, \citenamefont {Gaudi}, \citenamefont
  {Kerins}, \citenamefont {Rattenbury}, \citenamefont {Robin}, \citenamefont
  {Novati},\ and\ \citenamefont {Henderson}}]{Johnson_2020}%
  \BibitemOpen
  \bibfield  {author} {\bibinfo {author} {\bibfnamefont {S.~A.}\ \bibnamefont
  {Johnson}}, \bibinfo {author} {\bibfnamefont {M.}~\bibnamefont {Penny}},
  \bibinfo {author} {\bibfnamefont {B.~S.}\ \bibnamefont {Gaudi}}, \bibinfo
  {author} {\bibfnamefont {E.}~\bibnamefont {Kerins}}, \bibinfo {author}
  {\bibfnamefont {N.~J.}\ \bibnamefont {Rattenbury}}, \bibinfo {author}
  {\bibfnamefont {A.~C.}\ \bibnamefont {Robin}}, \bibinfo {author}
  {\bibfnamefont {S.~C.}\ \bibnamefont {Novati}}, \ and\ \bibinfo {author}
  {\bibfnamefont {C.~B.}\ \bibnamefont {Henderson}},\ }\href {\doibase
  10.3847/1538-3881/aba75b} {\bibfield  {journal} {\bibinfo  {journal} {The
  Astronomical Journal}\ }\textbf {\bibinfo {volume} {160}},\ \bibinfo {pages}
  {123} (\bibinfo {year} {2020})}\BibitemShut {NoStop}%
\bibitem [{\citenamefont {Benito}\ \emph {et~al.}(2021)\citenamefont {Benito},
  \citenamefont {Iocco},\ and\ \citenamefont {Cuoco}}]{Benito:2020lgu}%
  \BibitemOpen
  \bibfield  {author} {\bibinfo {author} {\bibfnamefont {M.}~\bibnamefont
  {Benito}}, \bibinfo {author} {\bibfnamefont {F.}~\bibnamefont {Iocco}}, \
  and\ \bibinfo {author} {\bibfnamefont {A.}~\bibnamefont {Cuoco}},\ }\href
  {\doibase 10.1016/j.dark.2021.100826} {\bibfield  {journal} {\bibinfo
  {journal} {Phys. Dark Univ.}\ }\textbf {\bibinfo {volume} {32}},\ \bibinfo
  {pages} {100826} (\bibinfo {year} {2021})},\ \Eprint
  {http://arxiv.org/abs/2009.13523} {arXiv:2009.13523 [astro-ph.GA]}
  \BibitemShut {NoStop}%
\bibitem [{\citenamefont {Batell}\ \emph {et~al.}(2010)\citenamefont {Batell},
  \citenamefont {Pospelov}, \citenamefont {Ritz},\ and\ \citenamefont
  {Shang}}]{Batell:2009zp}%
  \BibitemOpen
  \bibfield  {author} {\bibinfo {author} {\bibfnamefont {B.}~\bibnamefont
  {Batell}}, \bibinfo {author} {\bibfnamefont {M.}~\bibnamefont {Pospelov}},
  \bibinfo {author} {\bibfnamefont {A.}~\bibnamefont {Ritz}}, \ and\ \bibinfo
  {author} {\bibfnamefont {Y.}~\bibnamefont {Shang}},\ }\href {\doibase
  10.1103/PhysRevD.81.075004} {\bibfield  {journal} {\bibinfo  {journal} {Phys.
  Rev. D}\ }\textbf {\bibinfo {volume} {81}},\ \bibinfo {pages} {075004}
  (\bibinfo {year} {2010})},\ \Eprint {http://arxiv.org/abs/0910.1567}
  {arXiv:0910.1567 [hep-ph]} \BibitemShut {NoStop}%
\bibitem [{\citenamefont {Pospelov}\ \emph {et~al.}(2008)\citenamefont
  {Pospelov}, \citenamefont {Ritz},\ and\ \citenamefont
  {Voloshin}}]{Pospelov:2007mp}%
  \BibitemOpen
  \bibfield  {author} {\bibinfo {author} {\bibfnamefont {M.}~\bibnamefont
  {Pospelov}}, \bibinfo {author} {\bibfnamefont {A.}~\bibnamefont {Ritz}}, \
  and\ \bibinfo {author} {\bibfnamefont {M.~B.}\ \bibnamefont {Voloshin}},\
  }\href {\doibase 10.1016/j.physletb.2008.02.052} {\bibfield  {journal}
  {\bibinfo  {journal} {Phys. Lett. B}\ }\textbf {\bibinfo {volume} {662}},\
  \bibinfo {pages} {53} (\bibinfo {year} {2008})},\ \Eprint
  {http://arxiv.org/abs/0711.4866} {arXiv:0711.4866 [hep-ph]} \BibitemShut
  {NoStop}%
\bibitem [{\citenamefont {Pospelov}\ and\ \citenamefont
  {Ritz}(2009)}]{Pospelov:2008jd}%
  \BibitemOpen
  \bibfield  {author} {\bibinfo {author} {\bibfnamefont {M.}~\bibnamefont
  {Pospelov}}\ and\ \bibinfo {author} {\bibfnamefont {A.}~\bibnamefont
  {Ritz}},\ }\href {\doibase 10.1016/j.physletb.2008.12.012} {\bibfield
  {journal} {\bibinfo  {journal} {Phys. Lett. B}\ }\textbf {\bibinfo {volume}
  {671}},\ \bibinfo {pages} {391} (\bibinfo {year} {2009})},\ \Eprint
  {http://arxiv.org/abs/0810.1502} {arXiv:0810.1502 [hep-ph]} \BibitemShut
  {NoStop}%
\bibitem [{\citenamefont {Rothstein}\ \emph {et~al.}(2009)\citenamefont
  {Rothstein}, \citenamefont {Schwetz},\ and\ \citenamefont
  {Zupan}}]{Rothstein:2009pm}%
  \BibitemOpen
  \bibfield  {author} {\bibinfo {author} {\bibfnamefont {I.~Z.}\ \bibnamefont
  {Rothstein}}, \bibinfo {author} {\bibfnamefont {T.}~\bibnamefont {Schwetz}},
  \ and\ \bibinfo {author} {\bibfnamefont {J.}~\bibnamefont {Zupan}},\ }\href
  {\doibase 10.1088/1475-7516/2009/07/018} {\bibfield  {journal} {\bibinfo
  {journal} {JCAP}\ }\textbf {\bibinfo {volume} {07}},\ \bibinfo {pages} {018}
  (\bibinfo {year} {2009})},\ \Eprint {http://arxiv.org/abs/0903.3116}
  {arXiv:0903.3116 [astro-ph.HE]} \BibitemShut {NoStop}%
\bibitem [{\citenamefont {Chen}\ \emph {et~al.}(2009)\citenamefont {Chen},
  \citenamefont {Cline},\ and\ \citenamefont {Frey}}]{Chen:2009ab}%
  \BibitemOpen
  \bibfield  {author} {\bibinfo {author} {\bibfnamefont {F.}~\bibnamefont
  {Chen}}, \bibinfo {author} {\bibfnamefont {J.~M.}\ \bibnamefont {Cline}}, \
  and\ \bibinfo {author} {\bibfnamefont {A.~R.}\ \bibnamefont {Frey}},\ }\href
  {\doibase 10.1103/PhysRevD.80.083516} {\bibfield  {journal} {\bibinfo
  {journal} {Phys. Rev. D}\ }\textbf {\bibinfo {volume} {80}},\ \bibinfo
  {pages} {083516} (\bibinfo {year} {2009})},\ \Eprint
  {http://arxiv.org/abs/0907.4746} {arXiv:0907.4746 [hep-ph]} \BibitemShut
  {NoStop}%
\bibitem [{\citenamefont {Schuster}\ \emph
  {et~al.}(2010{\natexlab{a}})\citenamefont {Schuster}, \citenamefont {Toro},\
  and\ \citenamefont {Yavin}}]{Schuster:2009au}%
  \BibitemOpen
  \bibfield  {author} {\bibinfo {author} {\bibfnamefont {P.}~\bibnamefont
  {Schuster}}, \bibinfo {author} {\bibfnamefont {N.}~\bibnamefont {Toro}}, \
  and\ \bibinfo {author} {\bibfnamefont {I.}~\bibnamefont {Yavin}},\ }\href
  {\doibase 10.1103/PhysRevD.81.016002} {\bibfield  {journal} {\bibinfo
  {journal} {Phys. Rev. D}\ }\textbf {\bibinfo {volume} {81}},\ \bibinfo
  {pages} {016002} (\bibinfo {year} {2010}{\natexlab{a}})},\ \Eprint
  {http://arxiv.org/abs/0910.1602} {arXiv:0910.1602 [hep-ph]} \BibitemShut
  {NoStop}%
\bibitem [{\citenamefont {Schuster}\ \emph
  {et~al.}(2010{\natexlab{b}})\citenamefont {Schuster}, \citenamefont {Toro},
  \citenamefont {Weiner},\ and\ \citenamefont {Yavin}}]{Schuster:2009fc}%
  \BibitemOpen
  \bibfield  {author} {\bibinfo {author} {\bibfnamefont {P.}~\bibnamefont
  {Schuster}}, \bibinfo {author} {\bibfnamefont {N.}~\bibnamefont {Toro}},
  \bibinfo {author} {\bibfnamefont {N.}~\bibnamefont {Weiner}}, \ and\ \bibinfo
  {author} {\bibfnamefont {I.}~\bibnamefont {Yavin}},\ }\href {\doibase
  10.1103/PhysRevD.82.115012} {\bibfield  {journal} {\bibinfo  {journal} {Phys.
  Rev. D}\ }\textbf {\bibinfo {volume} {82}},\ \bibinfo {pages} {115012}
  (\bibinfo {year} {2010}{\natexlab{b}})},\ \Eprint
  {http://arxiv.org/abs/0910.1839} {arXiv:0910.1839 [hep-ph]} \BibitemShut
  {NoStop}%
\bibitem [{\citenamefont {Bell}\ and\ \citenamefont
  {Petraki}(2011)}]{Bell_2011}%
  \BibitemOpen
  \bibfield  {author} {\bibinfo {author} {\bibfnamefont {N.~F.}\ \bibnamefont
  {Bell}}\ and\ \bibinfo {author} {\bibfnamefont {K.}~\bibnamefont {Petraki}},\
  }\href {\doibase 10.1088/1475-7516/2011/04/003} {\bibfield  {journal}
  {\bibinfo  {journal} {Journal of Cosmology and Astroparticle Physics}\
  }\textbf {\bibinfo {volume} {2011}},\ \bibinfo {pages} {003–003} (\bibinfo
  {year} {2011})}\BibitemShut {NoStop}%
\bibitem [{\citenamefont {Feng}\ \emph
  {et~al.}(2016{\natexlab{a}})\citenamefont {Feng}, \citenamefont {Smolinsky},\
  and\ \citenamefont {Tanedo}}]{Feng:2015hja}%
  \BibitemOpen
  \bibfield  {author} {\bibinfo {author} {\bibfnamefont {J.~L.}\ \bibnamefont
  {Feng}}, \bibinfo {author} {\bibfnamefont {J.}~\bibnamefont {Smolinsky}}, \
  and\ \bibinfo {author} {\bibfnamefont {P.}~\bibnamefont {Tanedo}},\ }\href
  {\doibase 10.1103/PhysRevD.93.015014} {\bibfield  {journal} {\bibinfo
  {journal} {Phys. Rev. D}\ }\textbf {\bibinfo {volume} {93}},\ \bibinfo
  {pages} {015014} (\bibinfo {year} {2016}{\natexlab{a}})},\ \bibinfo {note}
  {[Erratum: Phys.Rev.D 96, 099901 (2017)]},\ \Eprint
  {http://arxiv.org/abs/1509.07525} {arXiv:1509.07525 [hep-ph]} \BibitemShut
  {NoStop}%
\bibitem [{\citenamefont {Kouvaris}\ and\ \citenamefont
  {Tinyakov}(2010)}]{Kouvaris:2010}%
  \BibitemOpen
  \bibfield  {author} {\bibinfo {author} {\bibfnamefont {C.}~\bibnamefont
  {Kouvaris}}\ and\ \bibinfo {author} {\bibfnamefont {P.}~\bibnamefont
  {Tinyakov}},\ }\href {\doibase 10.1103/PhysRevD.82.063531} {\bibfield
  {journal} {\bibinfo  {journal} {Phys. Rev. D}\ }\textbf {\bibinfo {volume}
  {82}},\ \bibinfo {pages} {063531} (\bibinfo {year} {2010})},\ \Eprint
  {http://arxiv.org/abs/1004.0586} {arXiv:1004.0586 [astro-ph.GA]} \BibitemShut
  {NoStop}%
\bibitem [{\citenamefont {Feng}\ \emph
  {et~al.}(2016{\natexlab{b}})\citenamefont {Feng}, \citenamefont {Smolinsky},\
  and\ \citenamefont {Tanedo}}]{Feng:2016ijc}%
  \BibitemOpen
  \bibfield  {author} {\bibinfo {author} {\bibfnamefont {J.~L.}\ \bibnamefont
  {Feng}}, \bibinfo {author} {\bibfnamefont {J.}~\bibnamefont {Smolinsky}}, \
  and\ \bibinfo {author} {\bibfnamefont {P.}~\bibnamefont {Tanedo}},\ }\href
  {\doibase 10.1103/PhysRevD.93.115036} {\bibfield  {journal} {\bibinfo
  {journal} {Phys. Rev. D}\ }\textbf {\bibinfo {volume} {93}},\ \bibinfo
  {pages} {115036} (\bibinfo {year} {2016}{\natexlab{b}})},\ \bibinfo {note}
  {[Erratum: Phys.Rev.D 96, 099903 (2017)]},\ \Eprint
  {http://arxiv.org/abs/1602.01465} {arXiv:1602.01465 [hep-ph]} \BibitemShut
  {NoStop}%
\bibitem [{\citenamefont {Allahverdi}\ \emph {et~al.}(2017)\citenamefont
  {Allahverdi}, \citenamefont {Gao}, \citenamefont {Knockel},\ and\
  \citenamefont {Shalgar}}]{Allahverdi:2016fvl}%
  \BibitemOpen
  \bibfield  {author} {\bibinfo {author} {\bibfnamefont {R.}~\bibnamefont
  {Allahverdi}}, \bibinfo {author} {\bibfnamefont {Y.}~\bibnamefont {Gao}},
  \bibinfo {author} {\bibfnamefont {B.}~\bibnamefont {Knockel}}, \ and\
  \bibinfo {author} {\bibfnamefont {S.}~\bibnamefont {Shalgar}},\ }\href
  {\doibase 10.1103/PhysRevD.95.075001} {\bibfield  {journal} {\bibinfo
  {journal} {Phys. Rev. D}\ }\textbf {\bibinfo {volume} {95}},\ \bibinfo
  {pages} {075001} (\bibinfo {year} {2017})},\ \Eprint
  {http://arxiv.org/abs/1612.03110} {arXiv:1612.03110 [hep-ph]} \BibitemShut
  {NoStop}%
\bibitem [{\citenamefont {Leane}\ \emph {et~al.}(2017)\citenamefont {Leane},
  \citenamefont {Ng},\ and\ \citenamefont {Beacom}}]{Leane:2017vag}%
  \BibitemOpen
  \bibfield  {author} {\bibinfo {author} {\bibfnamefont {R.~K.}\ \bibnamefont
  {Leane}}, \bibinfo {author} {\bibfnamefont {K.~C.~Y.}\ \bibnamefont {Ng}}, \
  and\ \bibinfo {author} {\bibfnamefont {J.~F.}\ \bibnamefont {Beacom}},\
  }\href {\doibase 10.1103/PhysRevD.95.123016} {\bibfield  {journal} {\bibinfo
  {journal} {Phys. Rev. D}\ }\textbf {\bibinfo {volume} {95}},\ \bibinfo
  {pages} {123016} (\bibinfo {year} {2017})},\ \Eprint
  {http://arxiv.org/abs/1703.04629} {arXiv:1703.04629 [astro-ph.HE]}
  \BibitemShut {NoStop}%
\bibitem [{\citenamefont {Arina}\ \emph {et~al.}(2017)\citenamefont {Arina},
  \citenamefont {Backovi\'c}, \citenamefont {Heisig},\ and\ \citenamefont
  {Lucente}}]{Arina:2017sng}%
  \BibitemOpen
  \bibfield  {author} {\bibinfo {author} {\bibfnamefont {C.}~\bibnamefont
  {Arina}}, \bibinfo {author} {\bibfnamefont {M.}~\bibnamefont {Backovi\'c}},
  \bibinfo {author} {\bibfnamefont {J.}~\bibnamefont {Heisig}}, \ and\ \bibinfo
  {author} {\bibfnamefont {M.}~\bibnamefont {Lucente}},\ }\href {\doibase
  10.1103/PhysRevD.96.063010} {\bibfield  {journal} {\bibinfo  {journal} {Phys.
  Rev. D}\ }\textbf {\bibinfo {volume} {96}},\ \bibinfo {pages} {063010}
  (\bibinfo {year} {2017})},\ \Eprint {http://arxiv.org/abs/1703.08087}
  {arXiv:1703.08087 [astro-ph.HE]} \BibitemShut {NoStop}%
\bibitem [{\citenamefont {Albert}\ \emph
  {et~al.}(2018{\natexlab{a}})\citenamefont {Albert} \emph
  {et~al.}}]{Albert:2018jwh}%
  \BibitemOpen
  \bibfield  {author} {\bibinfo {author} {\bibfnamefont {A.}~\bibnamefont
  {Albert}} \emph {et~al.} (\bibinfo {collaboration} {HAWC}),\ }\href {\doibase
  10.1103/PhysRevD.98.123012} {\bibfield  {journal} {\bibinfo  {journal} {Phys.
  Rev.}\ }\textbf {\bibinfo {volume} {D98}},\ \bibinfo {pages} {123012}
  (\bibinfo {year} {2018}{\natexlab{a}})},\ \Eprint
  {http://arxiv.org/abs/1808.05624} {arXiv:1808.05624 [hep-ph]} \BibitemShut
  {NoStop}%
\bibitem [{\citenamefont {Albert}\ \emph
  {et~al.}(2018{\natexlab{b}})\citenamefont {Albert} \emph
  {et~al.}}]{Albert:2018vcq}%
  \BibitemOpen
  \bibfield  {author} {\bibinfo {author} {\bibfnamefont {A.}~\bibnamefont
  {Albert}} \emph {et~al.} (\bibinfo {collaboration} {HAWC}),\ }\href {\doibase
  10.1103/PhysRevD.98.123011} {\bibfield  {journal} {\bibinfo  {journal} {Phys.
  Rev. D}\ }\textbf {\bibinfo {volume} {98}},\ \bibinfo {pages} {123011}
  (\bibinfo {year} {2018}{\natexlab{b}})},\ \Eprint
  {http://arxiv.org/abs/1808.05620} {arXiv:1808.05620 [astro-ph.HE]}
  \BibitemShut {NoStop}%
\bibitem [{\citenamefont {Nisa}\ \emph {et~al.}(2019)\citenamefont {Nisa},
  \citenamefont {Beacom}, \citenamefont {BenZvi}, \citenamefont {Leane},
  \citenamefont {Linden}, \citenamefont {Ng}, \citenamefont {Peter},\ and\
  \citenamefont {Zhou}}]{Nisa:2019mpb}%
  \BibitemOpen
  \bibfield  {author} {\bibinfo {author} {\bibfnamefont {M.~U.}\ \bibnamefont
  {Nisa}}, \bibinfo {author} {\bibfnamefont {J.~F.}\ \bibnamefont {Beacom}},
  \bibinfo {author} {\bibfnamefont {S.~Y.}\ \bibnamefont {BenZvi}}, \bibinfo
  {author} {\bibfnamefont {R.~K.}\ \bibnamefont {Leane}}, \bibinfo {author}
  {\bibfnamefont {T.}~\bibnamefont {Linden}}, \bibinfo {author} {\bibfnamefont
  {K.~C.~Y.}\ \bibnamefont {Ng}}, \bibinfo {author} {\bibfnamefont {A.~H.~G.}\
  \bibnamefont {Peter}}, \ and\ \bibinfo {author} {\bibfnamefont
  {B.}~\bibnamefont {Zhou}},\ }\href@noop {} {\  (\bibinfo {year} {2019})},\
  \Eprint {http://arxiv.org/abs/1903.06349} {arXiv:1903.06349 [astro-ph.HE]}
  \BibitemShut {NoStop}%
\bibitem [{\citenamefont {Niblaeus}\ \emph {et~al.}(2019)\citenamefont
  {Niblaeus}, \citenamefont {Beniwal},\ and\ \citenamefont
  {Edsjo}}]{Niblaeus:2019gjk}%
  \BibitemOpen
  \bibfield  {author} {\bibinfo {author} {\bibfnamefont {C.}~\bibnamefont
  {Niblaeus}}, \bibinfo {author} {\bibfnamefont {A.}~\bibnamefont {Beniwal}}, \
  and\ \bibinfo {author} {\bibfnamefont {J.}~\bibnamefont {Edsjo}},\ }\href
  {\doibase 10.1088/1475-7516/2019/11/011} {\bibfield  {journal} {\bibinfo
  {journal} {JCAP}\ }\textbf {\bibinfo {volume} {11}},\ \bibinfo {pages} {011}
  (\bibinfo {year} {2019})},\ \Eprint {http://arxiv.org/abs/1903.11363}
  {arXiv:1903.11363 [astro-ph.HE]} \BibitemShut {NoStop}%
\bibitem [{\citenamefont {Cuoco}\ \emph {et~al.}(2020)\citenamefont {Cuoco},
  \citenamefont {De~La Torre~Luque}, \citenamefont {Gargano}, \citenamefont
  {Gustafsson}, \citenamefont {Loparco}, \citenamefont {Mazziotta},\ and\
  \citenamefont {Serini}}]{Cuoco:2019mlb}%
  \BibitemOpen
  \bibfield  {author} {\bibinfo {author} {\bibfnamefont {A.}~\bibnamefont
  {Cuoco}}, \bibinfo {author} {\bibfnamefont {P.}~\bibnamefont {De~La
  Torre~Luque}}, \bibinfo {author} {\bibfnamefont {F.}~\bibnamefont {Gargano}},
  \bibinfo {author} {\bibfnamefont {M.}~\bibnamefont {Gustafsson}}, \bibinfo
  {author} {\bibfnamefont {F.}~\bibnamefont {Loparco}}, \bibinfo {author}
  {\bibfnamefont {M.}~\bibnamefont {Mazziotta}}, \ and\ \bibinfo {author}
  {\bibfnamefont {D.}~\bibnamefont {Serini}},\ }\href {\doibase
  10.1103/PhysRevD.101.022002} {\bibfield  {journal} {\bibinfo  {journal}
  {Phys. Rev. D}\ }\textbf {\bibinfo {volume} {101}},\ \bibinfo {pages}
  {022002} (\bibinfo {year} {2020})},\ \Eprint
  {http://arxiv.org/abs/1912.09373} {arXiv:1912.09373 [astro-ph.HE]}
  \BibitemShut {NoStop}%
\bibitem [{\citenamefont {Serini}\ \emph {et~al.}(2020)\citenamefont {Serini},
  \citenamefont {Loparco},\ and\ \citenamefont {Mazziotta}}]{Serini:2020yhb}%
  \BibitemOpen
  \bibfield  {author} {\bibinfo {author} {\bibfnamefont {D.}~\bibnamefont
  {Serini}}, \bibinfo {author} {\bibfnamefont {F.}~\bibnamefont {Loparco}}, \
  and\ \bibinfo {author} {\bibfnamefont {M.~N.}\ \bibnamefont {Mazziotta}}
  (\bibinfo {collaboration} {Fermi-LAT}),\ }\href {\doibase
  10.22323/1.358.0544} {\bibfield  {journal} {\bibinfo  {journal} {PoS}\
  }\textbf {\bibinfo {volume} {ICRC2019}},\ \bibinfo {pages} {544} (\bibinfo
  {year} {2020})}\BibitemShut {NoStop}%
\bibitem [{\citenamefont {Acevedo}\ \emph {et~al.}(2021)\citenamefont
  {Acevedo}, \citenamefont {Bramante}, \citenamefont {Goodman}, \citenamefont
  {Kopp},\ and\ \citenamefont {Opferkuch}}]{Acevedo:2020gro}%
  \BibitemOpen
  \bibfield  {author} {\bibinfo {author} {\bibfnamefont {J.~F.}\ \bibnamefont
  {Acevedo}}, \bibinfo {author} {\bibfnamefont {J.}~\bibnamefont {Bramante}},
  \bibinfo {author} {\bibfnamefont {A.}~\bibnamefont {Goodman}}, \bibinfo
  {author} {\bibfnamefont {J.}~\bibnamefont {Kopp}}, \ and\ \bibinfo {author}
  {\bibfnamefont {T.}~\bibnamefont {Opferkuch}},\ }\href {\doibase
  10.1088/1475-7516/2021/04/026} {\bibfield  {journal} {\bibinfo  {journal}
  {JCAP}\ }\textbf {\bibinfo {volume} {04}},\ \bibinfo {pages} {026} (\bibinfo
  {year} {2021})},\ \Eprint {http://arxiv.org/abs/2012.09176} {arXiv:2012.09176
  [hep-ph]} \BibitemShut {NoStop}%
\bibitem [{\citenamefont {Mazziotta}\ \emph {et~al.}(2020)\citenamefont
  {Mazziotta}, \citenamefont {Loparco}, \citenamefont {Serini}, \citenamefont
  {Cuoco}, \citenamefont {De~La Torre~Luque}, \citenamefont {Gargano},\ and\
  \citenamefont {Gustafsson}}]{Mazziotta:2020foa}%
  \BibitemOpen
  \bibfield  {author} {\bibinfo {author} {\bibfnamefont {M.}~\bibnamefont
  {Mazziotta}}, \bibinfo {author} {\bibfnamefont {F.}~\bibnamefont {Loparco}},
  \bibinfo {author} {\bibfnamefont {D.}~\bibnamefont {Serini}}, \bibinfo
  {author} {\bibfnamefont {A.}~\bibnamefont {Cuoco}}, \bibinfo {author}
  {\bibfnamefont {P.}~\bibnamefont {De~La Torre~Luque}}, \bibinfo {author}
  {\bibfnamefont {F.}~\bibnamefont {Gargano}}, \ and\ \bibinfo {author}
  {\bibfnamefont {M.}~\bibnamefont {Gustafsson}},\ }\href {\doibase
  10.1103/PhysRevD.102.022003} {\bibfield  {journal} {\bibinfo  {journal}
  {Phys. Rev. D}\ }\textbf {\bibinfo {volume} {102}},\ \bibinfo {pages}
  {022003} (\bibinfo {year} {2020})},\ \Eprint
  {http://arxiv.org/abs/2006.04114} {arXiv:2006.04114 [astro-ph.HE]}
  \BibitemShut {NoStop}%
\bibitem [{\citenamefont {Bell}\ \emph {et~al.}(2021)\citenamefont {Bell},
  \citenamefont {Dent},\ and\ \citenamefont {Sanderson}}]{Bell:2021pyy}%
  \BibitemOpen
  \bibfield  {author} {\bibinfo {author} {\bibfnamefont {N.~F.}\ \bibnamefont
  {Bell}}, \bibinfo {author} {\bibfnamefont {J.~B.}\ \bibnamefont {Dent}}, \
  and\ \bibinfo {author} {\bibfnamefont {I.~W.}\ \bibnamefont {Sanderson}},\
  }\href@noop {} {\  (\bibinfo {year} {2021})},\ \Eprint
  {http://arxiv.org/abs/2103.16794} {arXiv:2103.16794 [hep-ph]} \BibitemShut
  {NoStop}%
\bibitem [{\citenamefont {Leane}\ and\ \citenamefont
  {Linden}(2021)}]{Leane:2021tjj}%
  \BibitemOpen
  \bibfield  {author} {\bibinfo {author} {\bibfnamefont {R.~K.}\ \bibnamefont
  {Leane}}\ and\ \bibinfo {author} {\bibfnamefont {T.}~\bibnamefont {Linden}},\
  }\href@noop {} {\  (\bibinfo {year} {2021})},\ \Eprint
  {http://arxiv.org/abs/2104.02068} {arXiv:2104.02068 [astro-ph.HE]}
  \BibitemShut {NoStop}%
\bibitem [{\citenamefont {Li}\ and\ \citenamefont {Fan}(2022)}]{Li:2022wix}%
  \BibitemOpen
  \bibfield  {author} {\bibinfo {author} {\bibfnamefont {L.}~\bibnamefont
  {Li}}\ and\ \bibinfo {author} {\bibfnamefont {J.}~\bibnamefont {Fan}},\
  }\href {\doibase 10.1007/JHEP10(2022)186} {\bibfield  {journal} {\bibinfo
  {journal} {JHEP}\ }\textbf {\bibinfo {volume} {10}},\ \bibinfo {pages} {186}
  (\bibinfo {year} {2022})},\ \Eprint {http://arxiv.org/abs/2207.13709}
  {arXiv:2207.13709 [hep-ph]} \BibitemShut {NoStop}%
\bibitem [{\citenamefont {French}\ and\ \citenamefont
  {Sher}(2022)}]{French:2022ccb}%
  \BibitemOpen
  \bibfield  {author} {\bibinfo {author} {\bibfnamefont {G.~M.}\ \bibnamefont
  {French}}\ and\ \bibinfo {author} {\bibfnamefont {M.}~\bibnamefont {Sher}},\
  }\href {\doibase 10.1103/PhysRevD.106.115037} {\bibfield  {journal} {\bibinfo
   {journal} {Phys. Rev. D}\ }\textbf {\bibinfo {volume} {106}},\ \bibinfo
  {pages} {115037} (\bibinfo {year} {2022})},\ \Eprint
  {http://arxiv.org/abs/2210.04761} {arXiv:2210.04761 [hep-ph]} \BibitemShut
  {NoStop}%
\bibitem [{\citenamefont {Ray}(2023)}]{Ray:2023auh}%
  \BibitemOpen
  \bibfield  {author} {\bibinfo {author} {\bibfnamefont {A.}~\bibnamefont
  {Ray}},\ }\href {\doibase 10.1103/PhysRevD.107.083012} {\bibfield  {journal}
  {\bibinfo  {journal} {Phys. Rev. D}\ }\textbf {\bibinfo {volume} {107}},\
  \bibinfo {pages} {083012} (\bibinfo {year} {2023})},\ \Eprint
  {http://arxiv.org/abs/2301.03625} {arXiv:2301.03625 [hep-ph]} \BibitemShut
  {NoStop}%
\bibitem [{\citenamefont {Leane}\ and\ \citenamefont
  {Smirnov}(2021)}]{Leane:2020wob}%
  \BibitemOpen
  \bibfield  {author} {\bibinfo {author} {\bibfnamefont {R.~K.}\ \bibnamefont
  {Leane}}\ and\ \bibinfo {author} {\bibfnamefont {J.}~\bibnamefont
  {Smirnov}},\ }\href {\doibase 10.1103/PhysRevLett.126.161101} {\bibfield
  {journal} {\bibinfo  {journal} {Phys. Rev. Lett.}\ }\textbf {\bibinfo
  {volume} {126}},\ \bibinfo {pages} {161101} (\bibinfo {year} {2021})},\
  \Eprint {http://arxiv.org/abs/2010.00015} {arXiv:2010.00015 [hep-ph]}
  \BibitemShut {NoStop}%
\bibitem [{\citenamefont {Leane}\ \emph {et~al.}(2021)\citenamefont {Leane},
  \citenamefont {Linden}, \citenamefont {Mukhopadhyay},\ and\ \citenamefont
  {Toro}}]{Leane:2021ihh}%
  \BibitemOpen
  \bibfield  {author} {\bibinfo {author} {\bibfnamefont {R.~K.}\ \bibnamefont
  {Leane}}, \bibinfo {author} {\bibfnamefont {T.}~\bibnamefont {Linden}},
  \bibinfo {author} {\bibfnamefont {P.}~\bibnamefont {Mukhopadhyay}}, \ and\
  \bibinfo {author} {\bibfnamefont {N.}~\bibnamefont {Toro}},\ }\href {\doibase
  10.1103/PhysRevD.103.075030} {\bibfield  {journal} {\bibinfo  {journal}
  {Phys. Rev. D}\ }\textbf {\bibinfo {volume} {103}},\ \bibinfo {pages}
  {075030} (\bibinfo {year} {2021})},\ \Eprint
  {http://arxiv.org/abs/2101.12213} {arXiv:2101.12213 [astro-ph.HE]}
  \BibitemShut {NoStop}%
\bibitem [{\citenamefont {Mitra}(2004)}]{Mitra:2004fh}%
  \BibitemOpen
  \bibfield  {author} {\bibinfo {author} {\bibfnamefont {S.}~\bibnamefont
  {Mitra}},\ }\href {\doibase 10.1103/PhysRevD.70.103517} {\bibfield  {journal}
  {\bibinfo  {journal} {Phys. Rev. D}\ }\textbf {\bibinfo {volume} {70}},\
  \bibinfo {pages} {103517} (\bibinfo {year} {2004})},\ \Eprint
  {http://arxiv.org/abs/astro-ph/0408341} {arXiv:astro-ph/0408341} \BibitemShut
  {NoStop}%
\bibitem [{\citenamefont {Garani}\ \emph {et~al.}(2023)\citenamefont {Garani},
  \citenamefont {Raj},\ and\ \citenamefont {Reynoso-Cordova}}]{Garani:2023esk}%
  \BibitemOpen
  \bibfield  {author} {\bibinfo {author} {\bibfnamefont {R.}~\bibnamefont
  {Garani}}, \bibinfo {author} {\bibfnamefont {N.}~\bibnamefont {Raj}}, \ and\
  \bibinfo {author} {\bibfnamefont {J.}~\bibnamefont {Reynoso-Cordova}},\
  }\href {\doibase 10.1088/1475-7516/2023/07/038} {\bibfield  {journal}
  {\bibinfo  {journal} {JCAP}\ }\textbf {\bibinfo {volume} {07}},\ \bibinfo
  {pages} {038} (\bibinfo {year} {2023})},\ \Eprint
  {http://arxiv.org/abs/2303.18009} {arXiv:2303.18009 [astro-ph.HE]}
  \BibitemShut {NoStop}%
\bibitem [{\citenamefont {Goldman}\ and\ \citenamefont
  {Nussinov}(1989)}]{Goldman:1989nd}%
  \BibitemOpen
  \bibfield  {author} {\bibinfo {author} {\bibfnamefont {I.}~\bibnamefont
  {Goldman}}\ and\ \bibinfo {author} {\bibfnamefont {S.}~\bibnamefont
  {Nussinov}},\ }\href {\doibase 10.1103/PhysRevD.40.3221} {\bibfield
  {journal} {\bibinfo  {journal} {Phys. Rev. D}\ }\textbf {\bibinfo {volume}
  {40}},\ \bibinfo {pages} {3221} (\bibinfo {year} {1989})}\BibitemShut
  {NoStop}%
\bibitem [{\citenamefont {Gould}\ \emph {et~al.}(1990)\citenamefont {Gould},
  \citenamefont {Draine}, \citenamefont {Romani},\ and\ \citenamefont
  {Nussinov}}]{Gould:1989gw}%
  \BibitemOpen
  \bibfield  {author} {\bibinfo {author} {\bibfnamefont {A.}~\bibnamefont
  {Gould}}, \bibinfo {author} {\bibfnamefont {B.~T.}\ \bibnamefont {Draine}},
  \bibinfo {author} {\bibfnamefont {R.~W.}\ \bibnamefont {Romani}}, \ and\
  \bibinfo {author} {\bibfnamefont {S.}~\bibnamefont {Nussinov}},\ }\href
  {\doibase 10.1016/0370-2693(90)91745-W} {\bibfield  {journal} {\bibinfo
  {journal} {Phys. Lett. B}\ }\textbf {\bibinfo {volume} {238}},\ \bibinfo
  {pages} {337} (\bibinfo {year} {1990})}\BibitemShut {NoStop}%
\bibitem [{\citenamefont {Kouvaris}(2008)}]{Kouvaris:2007ay}%
  \BibitemOpen
  \bibfield  {author} {\bibinfo {author} {\bibfnamefont {C.}~\bibnamefont
  {Kouvaris}},\ }\href {\doibase 10.1103/PhysRevD.77.023006} {\bibfield
  {journal} {\bibinfo  {journal} {Phys. Rev. D}\ }\textbf {\bibinfo {volume}
  {77}},\ \bibinfo {pages} {023006} (\bibinfo {year} {2008})},\ \Eprint
  {http://arxiv.org/abs/0708.2362} {arXiv:0708.2362 [astro-ph]} \BibitemShut
  {NoStop}%
\bibitem [{\citenamefont {Bertone}\ and\ \citenamefont
  {Fairbairn}(2008)}]{Bertone:2007ae}%
  \BibitemOpen
  \bibfield  {author} {\bibinfo {author} {\bibfnamefont {G.}~\bibnamefont
  {Bertone}}\ and\ \bibinfo {author} {\bibfnamefont {M.}~\bibnamefont
  {Fairbairn}},\ }\href {\doibase 10.1103/PhysRevD.77.043515} {\bibfield
  {journal} {\bibinfo  {journal} {Phys. Rev. D}\ }\textbf {\bibinfo {volume}
  {77}},\ \bibinfo {pages} {043515} (\bibinfo {year} {2008})},\ \Eprint
  {http://arxiv.org/abs/0709.1485} {arXiv:0709.1485 [astro-ph]} \BibitemShut
  {NoStop}%
\bibitem [{\citenamefont {de~Lavallaz}\ and\ \citenamefont
  {Fairbairn}(2010)}]{deLavallaz:2010wp}%
  \BibitemOpen
  \bibfield  {author} {\bibinfo {author} {\bibfnamefont {A.}~\bibnamefont
  {de~Lavallaz}}\ and\ \bibinfo {author} {\bibfnamefont {M.}~\bibnamefont
  {Fairbairn}},\ }\href {\doibase 10.1103/PhysRevD.81.123521} {\bibfield
  {journal} {\bibinfo  {journal} {Phys. Rev. D}\ }\textbf {\bibinfo {volume}
  {81}},\ \bibinfo {pages} {123521} (\bibinfo {year} {2010})},\ \Eprint
  {http://arxiv.org/abs/1004.0629} {arXiv:1004.0629 [astro-ph.GA]} \BibitemShut
  {NoStop}%
\bibitem [{\citenamefont {McDermott}\ \emph {et~al.}(2012)\citenamefont
  {McDermott}, \citenamefont {Yu},\ and\ \citenamefont
  {Zurek}}]{McDermott:2011jp}%
  \BibitemOpen
  \bibfield  {author} {\bibinfo {author} {\bibfnamefont {S.~D.}\ \bibnamefont
  {McDermott}}, \bibinfo {author} {\bibfnamefont {H.-B.}\ \bibnamefont {Yu}}, \
  and\ \bibinfo {author} {\bibfnamefont {K.~M.}\ \bibnamefont {Zurek}},\ }\href
  {\doibase 10.1103/PhysRevD.85.023519} {\bibfield  {journal} {\bibinfo
  {journal} {Phys. Rev.}\ }\textbf {\bibinfo {volume} {D85}},\ \bibinfo {pages}
  {023519} (\bibinfo {year} {2012})},\ \Eprint {http://arxiv.org/abs/1103.5472}
  {arXiv:1103.5472 [hep-ph]} \BibitemShut {NoStop}%
\bibitem [{\citenamefont {Kouvaris}\ and\ \citenamefont
  {Tinyakov}(2011{\natexlab{a}})}]{Kouvaris:2011fi}%
  \BibitemOpen
  \bibfield  {author} {\bibinfo {author} {\bibfnamefont {C.}~\bibnamefont
  {Kouvaris}}\ and\ \bibinfo {author} {\bibfnamefont {P.}~\bibnamefont
  {Tinyakov}},\ }\href {\doibase 10.1103/PhysRevLett.107.091301} {\bibfield
  {journal} {\bibinfo  {journal} {Phys. Rev. Lett.}\ }\textbf {\bibinfo
  {volume} {107}},\ \bibinfo {pages} {091301} (\bibinfo {year}
  {2011}{\natexlab{a}})},\ \Eprint {http://arxiv.org/abs/1104.0382}
  {arXiv:1104.0382 [astro-ph.CO]} \BibitemShut {NoStop}%
\bibitem [{\citenamefont {Guver}\ \emph {et~al.}(2014)\citenamefont {Guver},
  \citenamefont {Erkoca}, \citenamefont {Hall~Reno},\ and\ \citenamefont
  {Sarcevic}}]{Guver:2012ba}%
  \BibitemOpen
  \bibfield  {author} {\bibinfo {author} {\bibfnamefont {T.}~\bibnamefont
  {Guver}}, \bibinfo {author} {\bibfnamefont {A.~E.}\ \bibnamefont {Erkoca}},
  \bibinfo {author} {\bibfnamefont {M.}~\bibnamefont {Hall~Reno}}, \ and\
  \bibinfo {author} {\bibfnamefont {I.}~\bibnamefont {Sarcevic}},\ }\href
  {\doibase 10.1088/1475-7516/2014/05/013} {\bibfield  {journal} {\bibinfo
  {journal} {JCAP}\ }\textbf {\bibinfo {volume} {1405}},\ \bibinfo {pages}
  {013} (\bibinfo {year} {2014})},\ \Eprint {http://arxiv.org/abs/1201.2400}
  {arXiv:1201.2400 [hep-ph]} \BibitemShut {NoStop}%
\bibitem [{\citenamefont {Bramante}\ \emph {et~al.}(2013)\citenamefont
  {Bramante}, \citenamefont {Fukushima},\ and\ \citenamefont
  {Kumar}}]{Bramante:2013hn}%
  \BibitemOpen
  \bibfield  {author} {\bibinfo {author} {\bibfnamefont {J.}~\bibnamefont
  {Bramante}}, \bibinfo {author} {\bibfnamefont {K.}~\bibnamefont {Fukushima}},
  \ and\ \bibinfo {author} {\bibfnamefont {J.}~\bibnamefont {Kumar}},\ }\href
  {\doibase 10.1103/PhysRevD.87.055012} {\bibfield  {journal} {\bibinfo
  {journal} {Phys. Rev.}\ }\textbf {\bibinfo {volume} {D87}},\ \bibinfo {pages}
  {055012} (\bibinfo {year} {2013})},\ \Eprint {http://arxiv.org/abs/1301.0036}
  {arXiv:1301.0036 [hep-ph]} \BibitemShut {NoStop}%
\bibitem [{\citenamefont {Bell}\ \emph {et~al.}(2013)\citenamefont {Bell},
  \citenamefont {Melatos},\ and\ \citenamefont {Petraki}}]{Bell:2013xk}%
  \BibitemOpen
  \bibfield  {author} {\bibinfo {author} {\bibfnamefont {N.~F.}\ \bibnamefont
  {Bell}}, \bibinfo {author} {\bibfnamefont {A.}~\bibnamefont {Melatos}}, \
  and\ \bibinfo {author} {\bibfnamefont {K.}~\bibnamefont {Petraki}},\ }\href
  {\doibase 10.1103/PhysRevD.87.123507} {\bibfield  {journal} {\bibinfo
  {journal} {Phys. Rev.}\ }\textbf {\bibinfo {volume} {D87}},\ \bibinfo {pages}
  {123507} (\bibinfo {year} {2013})},\ \Eprint {http://arxiv.org/abs/1301.6811}
  {arXiv:1301.6811 [hep-ph]} \BibitemShut {NoStop}%
\bibitem [{\citenamefont {Bramante}\ \emph {et~al.}(2014)\citenamefont
  {Bramante}, \citenamefont {Fukushima}, \citenamefont {Kumar},\ and\
  \citenamefont {Stopnitzky}}]{Bramante:2013nma}%
  \BibitemOpen
  \bibfield  {author} {\bibinfo {author} {\bibfnamefont {J.}~\bibnamefont
  {Bramante}}, \bibinfo {author} {\bibfnamefont {K.}~\bibnamefont {Fukushima}},
  \bibinfo {author} {\bibfnamefont {J.}~\bibnamefont {Kumar}}, \ and\ \bibinfo
  {author} {\bibfnamefont {E.}~\bibnamefont {Stopnitzky}},\ }\href {\doibase
  10.1103/PhysRevD.89.015010} {\bibfield  {journal} {\bibinfo  {journal} {Phys.
  Rev.}\ }\textbf {\bibinfo {volume} {D89}},\ \bibinfo {pages} {015010}
  (\bibinfo {year} {2014})},\ \Eprint {http://arxiv.org/abs/1310.3509}
  {arXiv:1310.3509 [hep-ph]} \BibitemShut {NoStop}%
\bibitem [{\citenamefont {Bertoni}\ \emph {et~al.}(2013)\citenamefont
  {Bertoni}, \citenamefont {Nelson},\ and\ \citenamefont
  {Reddy}}]{Bertoni:2013bsa}%
  \BibitemOpen
  \bibfield  {author} {\bibinfo {author} {\bibfnamefont {B.}~\bibnamefont
  {Bertoni}}, \bibinfo {author} {\bibfnamefont {A.~E.}\ \bibnamefont {Nelson}},
  \ and\ \bibinfo {author} {\bibfnamefont {S.}~\bibnamefont {Reddy}},\ }\href
  {\doibase 10.1103/PhysRevD.88.123505} {\bibfield  {journal} {\bibinfo
  {journal} {Phys. Rev. D}\ }\textbf {\bibinfo {volume} {88}},\ \bibinfo
  {pages} {123505} (\bibinfo {year} {2013})},\ \Eprint
  {http://arxiv.org/abs/1309.1721} {arXiv:1309.1721 [hep-ph]} \BibitemShut
  {NoStop}%
\bibitem [{\citenamefont {Kouvaris}\ and\ \citenamefont
  {Tinyakov}(2011{\natexlab{b}})}]{Kouvaris:2010jy}%
  \BibitemOpen
  \bibfield  {author} {\bibinfo {author} {\bibfnamefont {C.}~\bibnamefont
  {Kouvaris}}\ and\ \bibinfo {author} {\bibfnamefont {P.}~\bibnamefont
  {Tinyakov}},\ }\href {\doibase 10.1103/PhysRevD.83.083512} {\bibfield
  {journal} {\bibinfo  {journal} {Phys. Rev.}\ }\textbf {\bibinfo {volume}
  {D83}},\ \bibinfo {pages} {083512} (\bibinfo {year} {2011}{\natexlab{b}})},\
  \Eprint {http://arxiv.org/abs/1012.2039} {arXiv:1012.2039 [astro-ph.HE]}
  \BibitemShut {NoStop}%
\bibitem [{\citenamefont {McCullough}\ and\ \citenamefont
  {Fairbairn}(2010)}]{McCullough:2010ai}%
  \BibitemOpen
  \bibfield  {author} {\bibinfo {author} {\bibfnamefont {M.}~\bibnamefont
  {McCullough}}\ and\ \bibinfo {author} {\bibfnamefont {M.}~\bibnamefont
  {Fairbairn}},\ }\href {\doibase 10.1103/PhysRevD.81.083520} {\bibfield
  {journal} {\bibinfo  {journal} {Phys. Rev. D}\ }\textbf {\bibinfo {volume}
  {81}},\ \bibinfo {pages} {083520} (\bibinfo {year} {2010})},\ \Eprint
  {http://arxiv.org/abs/1001.2737} {arXiv:1001.2737 [hep-ph]} \BibitemShut
  {NoStop}%
\bibitem [{\citenamefont {Angeles Perez-Garcia}\ and\ \citenamefont
  {Silk}(2015)}]{Perez-Garcia:2014dra}%
  \BibitemOpen
  \bibfield  {author} {\bibinfo {author} {\bibfnamefont {M.}~\bibnamefont
  {Angeles Perez-Garcia}}\ and\ \bibinfo {author} {\bibfnamefont
  {J.}~\bibnamefont {Silk}},\ }\href {\doibase 10.1016/j.physletb.2015.03.026}
  {\bibfield  {journal} {\bibinfo  {journal} {Phys. Lett.}\ }\textbf {\bibinfo
  {volume} {B744}},\ \bibinfo {pages} {13} (\bibinfo {year} {2015})},\ \Eprint
  {http://arxiv.org/abs/1403.6111} {arXiv:1403.6111 [astro-ph.SR]} \BibitemShut
  {NoStop}%
\bibitem [{\citenamefont {Bramante}(2015)}]{Bramante:2015cua}%
  \BibitemOpen
  \bibfield  {author} {\bibinfo {author} {\bibfnamefont {J.}~\bibnamefont
  {Bramante}},\ }\href {\doibase 10.1103/PhysRevLett.115.141301} {\bibfield
  {journal} {\bibinfo  {journal} {Phys. Rev. Lett.}\ }\textbf {\bibinfo
  {volume} {115}},\ \bibinfo {pages} {141301} (\bibinfo {year} {2015})},\
  \Eprint {http://arxiv.org/abs/1505.07464} {arXiv:1505.07464 [hep-ph]}
  \BibitemShut {NoStop}%
\bibitem [{\citenamefont {Graham}\ \emph {et~al.}(2015)\citenamefont {Graham},
  \citenamefont {Rajendran},\ and\ \citenamefont {Varela}}]{Graham:2015apa}%
  \BibitemOpen
  \bibfield  {author} {\bibinfo {author} {\bibfnamefont {P.~W.}\ \bibnamefont
  {Graham}}, \bibinfo {author} {\bibfnamefont {S.}~\bibnamefont {Rajendran}}, \
  and\ \bibinfo {author} {\bibfnamefont {J.}~\bibnamefont {Varela}},\ }\href
  {\doibase 10.1103/PhysRevD.92.063007} {\bibfield  {journal} {\bibinfo
  {journal} {Phys. Rev.}\ }\textbf {\bibinfo {volume} {D92}},\ \bibinfo {pages}
  {063007} (\bibinfo {year} {2015})},\ \Eprint
  {http://arxiv.org/abs/1505.04444} {arXiv:1505.04444 [hep-ph]} \BibitemShut
  {NoStop}%
\bibitem [{\citenamefont {Cermeno}\ \emph {et~al.}(2016)\citenamefont
  {Cermeno}, \citenamefont {Perez-Garcia},\ and\ \citenamefont
  {Silk}}]{Cermeno:2016olb}%
  \BibitemOpen
  \bibfield  {author} {\bibinfo {author} {\bibfnamefont {M.}~\bibnamefont
  {Cermeno}}, \bibinfo {author} {\bibfnamefont {M.}~\bibnamefont
  {Perez-Garcia}}, \ and\ \bibinfo {author} {\bibfnamefont {J.}~\bibnamefont
  {Silk}},\ }\href {\doibase 10.1103/PhysRevD.94.063001} {\bibfield  {journal}
  {\bibinfo  {journal} {Phys. Rev.}\ }\textbf {\bibinfo {volume} {D94}},\
  \bibinfo {pages} {063001} (\bibinfo {year} {2016})},\ \Eprint
  {http://arxiv.org/abs/1607.06815} {arXiv:1607.06815 [astro-ph.HE]}
  \BibitemShut {NoStop}%
\bibitem [{\citenamefont {Graham}\ \emph {et~al.}(2018)\citenamefont {Graham},
  \citenamefont {Janish}, \citenamefont {Narayan}, \citenamefont {Rajendran},\
  and\ \citenamefont {Riggins}}]{Graham:2018efk}%
  \BibitemOpen
  \bibfield  {author} {\bibinfo {author} {\bibfnamefont {P.~W.}\ \bibnamefont
  {Graham}}, \bibinfo {author} {\bibfnamefont {R.}~\bibnamefont {Janish}},
  \bibinfo {author} {\bibfnamefont {V.}~\bibnamefont {Narayan}}, \bibinfo
  {author} {\bibfnamefont {S.}~\bibnamefont {Rajendran}}, \ and\ \bibinfo
  {author} {\bibfnamefont {P.}~\bibnamefont {Riggins}},\ }\href {\doibase
  10.1103/PhysRevD.98.115027} {\bibfield  {journal} {\bibinfo  {journal} {Phys.
  Rev.}\ }\textbf {\bibinfo {volume} {D98}},\ \bibinfo {pages} {115027}
  (\bibinfo {year} {2018})},\ \Eprint {http://arxiv.org/abs/1805.07381}
  {arXiv:1805.07381 [hep-ph]} \BibitemShut {NoStop}%
\bibitem [{\citenamefont {Acevedo}\ and\ \citenamefont
  {Bramante}(2019)}]{Acevedo:2019gre}%
  \BibitemOpen
  \bibfield  {author} {\bibinfo {author} {\bibfnamefont {J.~F.}\ \bibnamefont
  {Acevedo}}\ and\ \bibinfo {author} {\bibfnamefont {J.}~\bibnamefont
  {Bramante}},\ }\href {\doibase 10.1103/PhysRevD.100.043020} {\bibfield
  {journal} {\bibinfo  {journal} {Phys. Rev.}\ }\textbf {\bibinfo {volume}
  {D100}},\ \bibinfo {pages} {043020} (\bibinfo {year} {2019})},\ \Eprint
  {http://arxiv.org/abs/1904.11993} {arXiv:1904.11993 [hep-ph]} \BibitemShut
  {NoStop}%
\bibitem [{\citenamefont {Janish}\ \emph {et~al.}(2019)\citenamefont {Janish},
  \citenamefont {Narayan},\ and\ \citenamefont {Riggins}}]{Janish:2019nkk}%
  \BibitemOpen
  \bibfield  {author} {\bibinfo {author} {\bibfnamefont {R.}~\bibnamefont
  {Janish}}, \bibinfo {author} {\bibfnamefont {V.}~\bibnamefont {Narayan}}, \
  and\ \bibinfo {author} {\bibfnamefont {P.}~\bibnamefont {Riggins}},\ }\href
  {\doibase 10.1103/PhysRevD.100.035008} {\bibfield  {journal} {\bibinfo
  {journal} {Phys. Rev.}\ }\textbf {\bibinfo {volume} {D100}},\ \bibinfo
  {pages} {035008} (\bibinfo {year} {2019})},\ \Eprint
  {http://arxiv.org/abs/1905.00395} {arXiv:1905.00395 [hep-ph]} \BibitemShut
  {NoStop}%
\bibitem [{\citenamefont {Krall}\ and\ \citenamefont
  {Reece}(2018)}]{Krall:2017xij}%
  \BibitemOpen
  \bibfield  {author} {\bibinfo {author} {\bibfnamefont {R.}~\bibnamefont
  {Krall}}\ and\ \bibinfo {author} {\bibfnamefont {M.}~\bibnamefont {Reece}},\
  }\href {\doibase 10.1088/1674-1137/42/4/043105} {\bibfield  {journal}
  {\bibinfo  {journal} {Chin. Phys.}\ }\textbf {\bibinfo {volume} {C42}},\
  \bibinfo {pages} {043105} (\bibinfo {year} {2018})},\ \Eprint
  {http://arxiv.org/abs/1705.04843} {arXiv:1705.04843 [hep-ph]} \BibitemShut
  {NoStop}%
\bibitem [{\citenamefont {Baryakhtar}\ \emph {et~al.}(2017)\citenamefont
  {Baryakhtar}, \citenamefont {Bramante}, \citenamefont {Li}, \citenamefont
  {Linden},\ and\ \citenamefont {Raj}}]{Baryakhtar:2017dbj}%
  \BibitemOpen
  \bibfield  {author} {\bibinfo {author} {\bibfnamefont {M.}~\bibnamefont
  {Baryakhtar}}, \bibinfo {author} {\bibfnamefont {J.}~\bibnamefont
  {Bramante}}, \bibinfo {author} {\bibfnamefont {S.~W.}\ \bibnamefont {Li}},
  \bibinfo {author} {\bibfnamefont {T.}~\bibnamefont {Linden}}, \ and\ \bibinfo
  {author} {\bibfnamefont {N.}~\bibnamefont {Raj}},\ }\href {\doibase
  10.1103/PhysRevLett.119.131801} {\bibfield  {journal} {\bibinfo  {journal}
  {Phys. Rev. Lett.}\ }\textbf {\bibinfo {volume} {119}},\ \bibinfo {pages}
  {131801} (\bibinfo {year} {2017})},\ \Eprint
  {http://arxiv.org/abs/1704.01577} {arXiv:1704.01577 [hep-ph]} \BibitemShut
  {NoStop}%
\bibitem [{\citenamefont {Raj}\ \emph {et~al.}(2018)\citenamefont {Raj},
  \citenamefont {Tanedo},\ and\ \citenamefont {Yu}}]{Raj:2017wrv}%
  \BibitemOpen
  \bibfield  {author} {\bibinfo {author} {\bibfnamefont {N.}~\bibnamefont
  {Raj}}, \bibinfo {author} {\bibfnamefont {P.}~\bibnamefont {Tanedo}}, \ and\
  \bibinfo {author} {\bibfnamefont {H.-B.}\ \bibnamefont {Yu}},\ }\href
  {\doibase 10.1103/PhysRevD.97.043006} {\bibfield  {journal} {\bibinfo
  {journal} {Phys. Rev.}\ }\textbf {\bibinfo {volume} {D97}},\ \bibinfo {pages}
  {043006} (\bibinfo {year} {2018})},\ \Eprint
  {http://arxiv.org/abs/1707.09442} {arXiv:1707.09442 [hep-ph]} \BibitemShut
  {NoStop}%
\bibitem [{\citenamefont {Bell}\ \emph {et~al.}(2018)\citenamefont {Bell},
  \citenamefont {Busoni},\ and\ \citenamefont {Robles}}]{Bell:2018pkk}%
  \BibitemOpen
  \bibfield  {author} {\bibinfo {author} {\bibfnamefont {N.~F.}\ \bibnamefont
  {Bell}}, \bibinfo {author} {\bibfnamefont {G.}~\bibnamefont {Busoni}}, \ and\
  \bibinfo {author} {\bibfnamefont {S.}~\bibnamefont {Robles}},\ }\href
  {\doibase 10.1088/1475-7516/2018/09/018} {\bibfield  {journal} {\bibinfo
  {journal} {JCAP}\ }\textbf {\bibinfo {volume} {1809}},\ \bibinfo {pages}
  {018} (\bibinfo {year} {2018})},\ \Eprint {http://arxiv.org/abs/1807.02840}
  {arXiv:1807.02840 [hep-ph]} \BibitemShut {NoStop}%
\bibitem [{\citenamefont {Chen}\ and\ \citenamefont
  {Lin}(2018)}]{Chen:2018ohx}%
  \BibitemOpen
  \bibfield  {author} {\bibinfo {author} {\bibfnamefont {C.-S.}\ \bibnamefont
  {Chen}}\ and\ \bibinfo {author} {\bibfnamefont {Y.-H.}\ \bibnamefont {Lin}},\
  }\href {\doibase 10.1007/JHEP08(2018)069} {\bibfield  {journal} {\bibinfo
  {journal} {JHEP}\ }\textbf {\bibinfo {volume} {08}},\ \bibinfo {pages} {069}
  (\bibinfo {year} {2018})},\ \Eprint {http://arxiv.org/abs/1804.03409}
  {arXiv:1804.03409 [hep-ph]} \BibitemShut {NoStop}%
\bibitem [{\citenamefont {Dasgupta}\ \emph {et~al.}(2019)\citenamefont
  {Dasgupta}, \citenamefont {Gupta},\ and\ \citenamefont
  {Ray}}]{Dasgupta:2019juq}%
  \BibitemOpen
  \bibfield  {author} {\bibinfo {author} {\bibfnamefont {B.}~\bibnamefont
  {Dasgupta}}, \bibinfo {author} {\bibfnamefont {A.}~\bibnamefont {Gupta}}, \
  and\ \bibinfo {author} {\bibfnamefont {A.}~\bibnamefont {Ray}},\ }\href
  {\doibase 10.1088/1475-7516/2019/08/018} {\bibfield  {journal} {\bibinfo
  {journal} {JCAP}\ }\textbf {\bibinfo {volume} {08}},\ \bibinfo {pages} {018}
  (\bibinfo {year} {2019})},\ \Eprint {http://arxiv.org/abs/1906.04204}
  {arXiv:1906.04204 [hep-ph]} \BibitemShut {NoStop}%
\bibitem [{\citenamefont {Hamaguchi}\ \emph {et~al.}(2019)\citenamefont
  {Hamaguchi}, \citenamefont {Nagata},\ and\ \citenamefont
  {Yanagi}}]{Hamaguchi:2019oev}%
  \BibitemOpen
  \bibfield  {author} {\bibinfo {author} {\bibfnamefont {K.}~\bibnamefont
  {Hamaguchi}}, \bibinfo {author} {\bibfnamefont {N.}~\bibnamefont {Nagata}}, \
  and\ \bibinfo {author} {\bibfnamefont {K.}~\bibnamefont {Yanagi}},\ }\href
  {\doibase 10.1016/j.physletb.2019.06.060} {\bibfield  {journal} {\bibinfo
  {journal} {Phys. Lett.}\ }\textbf {\bibinfo {volume} {B795}},\ \bibinfo
  {pages} {484} (\bibinfo {year} {2019})},\ \Eprint
  {http://arxiv.org/abs/1905.02991} {arXiv:1905.02991 [hep-ph]} \BibitemShut
  {NoStop}%
\bibitem [{\citenamefont {Camargo}\ \emph {et~al.}(2019)\citenamefont
  {Camargo}, \citenamefont {Queiroz},\ and\ \citenamefont
  {Sturani}}]{Camargo:2019wou}%
  \BibitemOpen
  \bibfield  {author} {\bibinfo {author} {\bibfnamefont {D.~A.}\ \bibnamefont
  {Camargo}}, \bibinfo {author} {\bibfnamefont {F.~S.}\ \bibnamefont
  {Queiroz}}, \ and\ \bibinfo {author} {\bibfnamefont {R.}~\bibnamefont
  {Sturani}},\ }\href {\doibase 10.1088/1475-7516/2019/09/051} {\bibfield
  {journal} {\bibinfo  {journal} {JCAP}\ }\textbf {\bibinfo {volume} {1909}},\
  \bibinfo {pages} {051} (\bibinfo {year} {2019})},\ \Eprint
  {http://arxiv.org/abs/1901.05474} {arXiv:1901.05474 [hep-ph]} \BibitemShut
  {NoStop}%
\bibitem [{\citenamefont {Bell}\ \emph {et~al.}(2019)\citenamefont {Bell},
  \citenamefont {Busoni},\ and\ \citenamefont {Robles}}]{Bell:2019pyc}%
  \BibitemOpen
  \bibfield  {author} {\bibinfo {author} {\bibfnamefont {N.~F.}\ \bibnamefont
  {Bell}}, \bibinfo {author} {\bibfnamefont {G.}~\bibnamefont {Busoni}}, \ and\
  \bibinfo {author} {\bibfnamefont {S.}~\bibnamefont {Robles}},\ }\href
  {\doibase 10.1088/1475-7516/2019/06/054} {\bibfield  {journal} {\bibinfo
  {journal} {JCAP}\ }\textbf {\bibinfo {volume} {1906}},\ \bibinfo {pages}
  {054} (\bibinfo {year} {2019})},\ \Eprint {http://arxiv.org/abs/1904.09803}
  {arXiv:1904.09803 [hep-ph]} \BibitemShut {NoStop}%
\bibitem [{\citenamefont {Acevedo}\ \emph {et~al.}(2020)\citenamefont
  {Acevedo}, \citenamefont {Bramante}, \citenamefont {Leane},\ and\
  \citenamefont {Raj}}]{Acevedo:2019agu}%
  \BibitemOpen
  \bibfield  {author} {\bibinfo {author} {\bibfnamefont {J.~F.}\ \bibnamefont
  {Acevedo}}, \bibinfo {author} {\bibfnamefont {J.}~\bibnamefont {Bramante}},
  \bibinfo {author} {\bibfnamefont {R.~K.}\ \bibnamefont {Leane}}, \ and\
  \bibinfo {author} {\bibfnamefont {N.}~\bibnamefont {Raj}},\ }\href {\doibase
  10.1088/1475-7516/2020/03/038} {\bibfield  {journal} {\bibinfo  {journal}
  {JCAP}\ }\textbf {\bibinfo {volume} {03}},\ \bibinfo {pages} {038} (\bibinfo
  {year} {2020})},\ \Eprint {http://arxiv.org/abs/1911.06334} {arXiv:1911.06334
  [hep-ph]} \BibitemShut {NoStop}%
\bibitem [{\citenamefont {Joglekar}\ \emph {et~al.}(2019)\citenamefont
  {Joglekar}, \citenamefont {Raj}, \citenamefont {Tanedo},\ and\ \citenamefont
  {Yu}}]{Joglekar:2019vzy}%
  \BibitemOpen
  \bibfield  {author} {\bibinfo {author} {\bibfnamefont {A.}~\bibnamefont
  {Joglekar}}, \bibinfo {author} {\bibfnamefont {N.}~\bibnamefont {Raj}},
  \bibinfo {author} {\bibfnamefont {P.}~\bibnamefont {Tanedo}}, \ and\ \bibinfo
  {author} {\bibfnamefont {H.-B.}\ \bibnamefont {Yu}},\ }\href@noop {} {\
  (\bibinfo {year} {2019})},\ \Eprint {http://arxiv.org/abs/1911.13293}
  {arXiv:1911.13293 [hep-ph]} \BibitemShut {NoStop}%
\bibitem [{\citenamefont {Joglekar}\ \emph {et~al.}(2020)\citenamefont
  {Joglekar}, \citenamefont {Raj}, \citenamefont {Tanedo},\ and\ \citenamefont
  {Yu}}]{Joglekar:2020liw}%
  \BibitemOpen
  \bibfield  {author} {\bibinfo {author} {\bibfnamefont {A.}~\bibnamefont
  {Joglekar}}, \bibinfo {author} {\bibfnamefont {N.}~\bibnamefont {Raj}},
  \bibinfo {author} {\bibfnamefont {P.}~\bibnamefont {Tanedo}}, \ and\ \bibinfo
  {author} {\bibfnamefont {H.-B.}\ \bibnamefont {Yu}},\ }\href@noop {} {\
  (\bibinfo {year} {2020})},\ \Eprint {http://arxiv.org/abs/2004.09539}
  {arXiv:2004.09539 [hep-ph]} \BibitemShut {NoStop}%
\bibitem [{\citenamefont {Bell}\ \emph {et~al.}(2020)\citenamefont {Bell},
  \citenamefont {Busoni}, \citenamefont {Robles},\ and\ \citenamefont
  {Virgato}}]{Bell:2020jou}%
  \BibitemOpen
  \bibfield  {author} {\bibinfo {author} {\bibfnamefont {N.~F.}\ \bibnamefont
  {Bell}}, \bibinfo {author} {\bibfnamefont {G.}~\bibnamefont {Busoni}},
  \bibinfo {author} {\bibfnamefont {S.}~\bibnamefont {Robles}}, \ and\ \bibinfo
  {author} {\bibfnamefont {M.}~\bibnamefont {Virgato}},\ }\href@noop {} {\
  (\bibinfo {year} {2020})},\ \Eprint {http://arxiv.org/abs/2004.14888}
  {arXiv:2004.14888 [hep-ph]} \BibitemShut {NoStop}%
\bibitem [{\citenamefont {Garani}\ \emph {et~al.}(2020)\citenamefont {Garani},
  \citenamefont {Gupta},\ and\ \citenamefont {Raj}}]{Garani:2020wge}%
  \BibitemOpen
  \bibfield  {author} {\bibinfo {author} {\bibfnamefont {R.}~\bibnamefont
  {Garani}}, \bibinfo {author} {\bibfnamefont {A.}~\bibnamefont {Gupta}}, \
  and\ \bibinfo {author} {\bibfnamefont {N.}~\bibnamefont {Raj}},\ }\href@noop
  {} {\  (\bibinfo {year} {2020})},\ \Eprint {http://arxiv.org/abs/2009.10728}
  {arXiv:2009.10728 [hep-ph]} \BibitemShut {NoStop}%
\bibitem [{\citenamefont {Bramante}\ \emph
  {et~al.}(2022{\natexlab{a}})\citenamefont {Bramante}, \citenamefont
  {Kavanagh},\ and\ \citenamefont {Raj}}]{Bramante:2021dyx}%
  \BibitemOpen
  \bibfield  {author} {\bibinfo {author} {\bibfnamefont {J.}~\bibnamefont
  {Bramante}}, \bibinfo {author} {\bibfnamefont {B.~J.}\ \bibnamefont
  {Kavanagh}}, \ and\ \bibinfo {author} {\bibfnamefont {N.}~\bibnamefont
  {Raj}},\ }\href {\doibase 10.1103/PhysRevLett.128.231801} {\bibfield
  {journal} {\bibinfo  {journal} {Phys. Rev. Lett.}\ }\textbf {\bibinfo
  {volume} {128}},\ \bibinfo {pages} {231801} (\bibinfo {year}
  {2022}{\natexlab{a}})},\ \Eprint {http://arxiv.org/abs/2109.04582}
  {arXiv:2109.04582 [hep-ph]} \BibitemShut {NoStop}%
\bibitem [{\citenamefont {Coffey}\ \emph {et~al.}(2022)\citenamefont {Coffey},
  \citenamefont {McKeen}, \citenamefont {Morrissey},\ and\ \citenamefont
  {Raj}}]{Coffey:2022eav}%
  \BibitemOpen
  \bibfield  {author} {\bibinfo {author} {\bibfnamefont {J.}~\bibnamefont
  {Coffey}}, \bibinfo {author} {\bibfnamefont {D.}~\bibnamefont {McKeen}},
  \bibinfo {author} {\bibfnamefont {D.~E.}\ \bibnamefont {Morrissey}}, \ and\
  \bibinfo {author} {\bibfnamefont {N.}~\bibnamefont {Raj}},\ }\href {\doibase
  10.1103/PhysRevD.106.115019} {\bibfield  {journal} {\bibinfo  {journal}
  {Phys. Rev. D}\ }\textbf {\bibinfo {volume} {106}},\ \bibinfo {pages}
  {115019} (\bibinfo {year} {2022})},\ \Eprint
  {http://arxiv.org/abs/2207.02221} {arXiv:2207.02221 [hep-ph]} \BibitemShut
  {NoStop}%
\bibitem [{\citenamefont {Freese}\ \emph {et~al.}(2009)\citenamefont {Freese},
  \citenamefont {Gondolo}, \citenamefont {Sellwood},\ and\ \citenamefont
  {Spolyar}}]{Freese:2008hb}%
  \BibitemOpen
  \bibfield  {author} {\bibinfo {author} {\bibfnamefont {K.}~\bibnamefont
  {Freese}}, \bibinfo {author} {\bibfnamefont {P.}~\bibnamefont {Gondolo}},
  \bibinfo {author} {\bibfnamefont {J.~A.}\ \bibnamefont {Sellwood}}, \ and\
  \bibinfo {author} {\bibfnamefont {D.}~\bibnamefont {Spolyar}},\ }\href
  {\doibase 10.1088/0004-637X/693/2/1563} {\bibfield  {journal} {\bibinfo
  {journal} {Astrophys. J.}\ }\textbf {\bibinfo {volume} {693}},\ \bibinfo
  {pages} {1563} (\bibinfo {year} {2009})},\ \Eprint
  {http://arxiv.org/abs/0805.3540} {arXiv:0805.3540 [astro-ph]} \BibitemShut
  {NoStop}%
\bibitem [{\citenamefont {Taoso}\ \emph {et~al.}(2008)\citenamefont {Taoso},
  \citenamefont {Bertone}, \citenamefont {Meynet},\ and\ \citenamefont
  {Ekstrom}}]{Taoso:2008kw}%
  \BibitemOpen
  \bibfield  {author} {\bibinfo {author} {\bibfnamefont {M.}~\bibnamefont
  {Taoso}}, \bibinfo {author} {\bibfnamefont {G.}~\bibnamefont {Bertone}},
  \bibinfo {author} {\bibfnamefont {G.}~\bibnamefont {Meynet}}, \ and\ \bibinfo
  {author} {\bibfnamefont {S.}~\bibnamefont {Ekstrom}},\ }\href {\doibase
  10.1103/PhysRevD.78.123510} {\bibfield  {journal} {\bibinfo  {journal} {Phys.
  Rev. D}\ }\textbf {\bibinfo {volume} {78}},\ \bibinfo {pages} {123510}
  (\bibinfo {year} {2008})},\ \Eprint {http://arxiv.org/abs/0806.2681}
  {arXiv:0806.2681 [astro-ph]} \BibitemShut {NoStop}%
\bibitem [{\citenamefont {Ilie}\ \emph
  {et~al.}(2020{\natexlab{a}})\citenamefont {Ilie}, \citenamefont {Levy},
  \citenamefont {Pilawa},\ and\ \citenamefont {Zhang}}]{Ilie:2020iup}%
  \BibitemOpen
  \bibfield  {author} {\bibinfo {author} {\bibfnamefont {C.}~\bibnamefont
  {Ilie}}, \bibinfo {author} {\bibfnamefont {C.}~\bibnamefont {Levy}}, \bibinfo
  {author} {\bibfnamefont {J.}~\bibnamefont {Pilawa}}, \ and\ \bibinfo {author}
  {\bibfnamefont {S.}~\bibnamefont {Zhang}},\ }\href@noop {} {\  (\bibinfo
  {year} {2020}{\natexlab{a}})},\ \Eprint {http://arxiv.org/abs/2009.11478}
  {arXiv:2009.11478 [astro-ph.CO]} \BibitemShut {NoStop}%
\bibitem [{\citenamefont {Ilie}\ \emph
  {et~al.}(2020{\natexlab{b}})\citenamefont {Ilie}, \citenamefont {Levy},
  \citenamefont {Pilawa},\ and\ \citenamefont {Zhang}}]{Ilie:2020nzp}%
  \BibitemOpen
  \bibfield  {author} {\bibinfo {author} {\bibfnamefont {C.}~\bibnamefont
  {Ilie}}, \bibinfo {author} {\bibfnamefont {C.}~\bibnamefont {Levy}}, \bibinfo
  {author} {\bibfnamefont {J.}~\bibnamefont {Pilawa}}, \ and\ \bibinfo {author}
  {\bibfnamefont {S.}~\bibnamefont {Zhang}},\ }\href@noop {} {\  (\bibinfo
  {year} {2020}{\natexlab{b}})},\ \Eprint {http://arxiv.org/abs/2009.11474}
  {arXiv:2009.11474 [astro-ph.CO]} \BibitemShut {NoStop}%
\bibitem [{\citenamefont {Lopes}\ and\ \citenamefont
  {Lopes}(2021)}]{Lopes:2021jcy}%
  \BibitemOpen
  \bibfield  {author} {\bibinfo {author} {\bibfnamefont {J.}~\bibnamefont
  {Lopes}}\ and\ \bibinfo {author} {\bibfnamefont {I.}~\bibnamefont {Lopes}},\
  }\href {\doibase 10.1051/0004-6361/202140750} {\bibfield  {journal} {\bibinfo
   {journal} {Astron. Astrophys.}\ }\textbf {\bibinfo {volume} {651}},\
  \bibinfo {pages} {A101} (\bibinfo {year} {2021})},\ \Eprint
  {http://arxiv.org/abs/2107.13885} {arXiv:2107.13885 [astro-ph.SR]}
  \BibitemShut {NoStop}%
\bibitem [{\citenamefont {Ellis}(2021)}]{Ellis:2021ztw}%
  \BibitemOpen
  \bibfield  {author} {\bibinfo {author} {\bibfnamefont {S.~A.~R.}\
  \bibnamefont {Ellis}},\ }\href@noop {} {\  (\bibinfo {year} {2021})},\
  \Eprint {http://arxiv.org/abs/2111.02414} {arXiv:2111.02414 [astro-ph.CO]}
  \BibitemShut {NoStop}%
\bibitem [{\citenamefont {Croon}\ and\ \citenamefont
  {Sakstein}(2023)}]{DCandJS}%
  \BibitemOpen
  \bibfield  {author} {\bibinfo {author} {\bibfnamefont {D.}~\bibnamefont
  {Croon}}\ and\ \bibinfo {author} {\bibfnamefont {J.}~\bibnamefont
  {Sakstein}},\ }\href@noop {} {\  (\bibinfo {year} {2023})},\ \Eprint
  {http://arxiv.org/abs/2309.XXXXX} {arXiv:2309.XXXXX [hep-ph]} \BibitemShut
  {NoStop}%
\bibitem [{\citenamefont {Bramante}\ \emph
  {et~al.}(2022{\natexlab{b}})\citenamefont {Bramante}, \citenamefont {Kumar},
  \citenamefont {Mohlabeng}, \citenamefont {Raj},\ and\ \citenamefont
  {Song}}]{Bramante:2022pmn}%
  \BibitemOpen
  \bibfield  {author} {\bibinfo {author} {\bibfnamefont {J.}~\bibnamefont
  {Bramante}}, \bibinfo {author} {\bibfnamefont {J.}~\bibnamefont {Kumar}},
  \bibinfo {author} {\bibfnamefont {G.}~\bibnamefont {Mohlabeng}}, \bibinfo
  {author} {\bibfnamefont {N.}~\bibnamefont {Raj}}, \ and\ \bibinfo {author}
  {\bibfnamefont {N.}~\bibnamefont {Song}},\ }\href@noop {} {\  (\bibinfo
  {year} {2022}{\natexlab{b}})},\ \Eprint {http://arxiv.org/abs/2210.01812}
  {arXiv:2210.01812 [hep-ph]} \BibitemShut {NoStop}%
\bibitem [{\citenamefont {Bramante}\ and\ \citenamefont
  {Raj}(2023)}]{Bramante:2023djs}%
  \BibitemOpen
  \bibfield  {author} {\bibinfo {author} {\bibfnamefont {J.}~\bibnamefont
  {Bramante}}\ and\ \bibinfo {author} {\bibfnamefont {N.}~\bibnamefont {Raj}},\
  }\href@noop {} {\  (\bibinfo {year} {2023})},\ \Eprint
  {http://arxiv.org/abs/2307.14435} {arXiv:2307.14435 [hep-ph]} \BibitemShut
  {NoStop}%
\bibitem [{\citenamefont {{Hubickyj}}\ \emph {et~al.}(2005)\citenamefont
  {{Hubickyj}}, \citenamefont {{Bodenheimer}},\ and\ \citenamefont
  {{Lissauer}}}]{2005Icar..179..415H}%
  \BibitemOpen
  \bibfield  {author} {\bibinfo {author} {\bibfnamefont {O.}~\bibnamefont
  {{Hubickyj}}}, \bibinfo {author} {\bibfnamefont {P.}~\bibnamefont
  {{Bodenheimer}}}, \ and\ \bibinfo {author} {\bibfnamefont {J.~J.}\
  \bibnamefont {{Lissauer}}},\ }\href {\doibase 10.1016/j.icarus.2005.06.021}
  {\bibfield  {journal} {\bibinfo  {journal} {\icarus}\ }\textbf {\bibinfo
  {volume} {179}},\ \bibinfo {pages} {415} (\bibinfo {year}
  {2005})}\BibitemShut {NoStop}%
\bibitem [{\citenamefont {Pollack}\ \emph {et~al.}(1996)\citenamefont
  {Pollack}, \citenamefont {Hubickyj}, \citenamefont {Bodenheimer},
  \citenamefont {Lissauer}, \citenamefont {Podolak},\ and\ \citenamefont
  {Greenzweig}}]{pollack1996formation}%
  \BibitemOpen
  \bibfield  {author} {\bibinfo {author} {\bibfnamefont {J.~B.}\ \bibnamefont
  {Pollack}}, \bibinfo {author} {\bibfnamefont {O.}~\bibnamefont {Hubickyj}},
  \bibinfo {author} {\bibfnamefont {P.}~\bibnamefont {Bodenheimer}}, \bibinfo
  {author} {\bibfnamefont {J.~J.}\ \bibnamefont {Lissauer}}, \bibinfo {author}
  {\bibfnamefont {M.}~\bibnamefont {Podolak}}, \ and\ \bibinfo {author}
  {\bibfnamefont {Y.}~\bibnamefont {Greenzweig}},\ }\href@noop {} {\bibfield
  {journal} {\bibinfo  {journal} {icarus}\ }\textbf {\bibinfo {volume} {124}},\
  \bibinfo {pages} {62} (\bibinfo {year} {1996})}\BibitemShut {NoStop}%
\bibitem [{\citenamefont {{Adams}}\ and\ \citenamefont
  {{Batygin}}(2022)}]{2022DPS....5410202A}%
  \BibitemOpen
  \bibfield  {author} {\bibinfo {author} {\bibfnamefont {F.}~\bibnamefont
  {{Adams}}}\ and\ \bibinfo {author} {\bibfnamefont {K.}~\bibnamefont
  {{Batygin}}},\ }in\ \href@noop {} {\emph {\bibinfo {booktitle} {AAS/Division
  for Planetary Sciences Meeting Abstracts}}},\ \bibinfo {series} {AAS/Division
  for Planetary Sciences Meeting Abstracts}, Vol.~\bibinfo {volume} {54}\
  (\bibinfo {year} {2022})\ p.\ \bibinfo {pages} {102.02}\BibitemShut {NoStop}%
\bibitem [{\citenamefont {{D'Angelo}}\ \emph {et~al.}(2021)\citenamefont
  {{D'Angelo}}, \citenamefont {{Weidenschilling}}, \citenamefont {{Lissauer}},\
  and\ \citenamefont {{Bodenheimer}}}]{2021Icar..35514087D}%
  \BibitemOpen
  \bibfield  {author} {\bibinfo {author} {\bibfnamefont {G.}~\bibnamefont
  {{D'Angelo}}}, \bibinfo {author} {\bibfnamefont {S.~J.}\ \bibnamefont
  {{Weidenschilling}}}, \bibinfo {author} {\bibfnamefont {J.~J.}\ \bibnamefont
  {{Lissauer}}}, \ and\ \bibinfo {author} {\bibfnamefont {P.}~\bibnamefont
  {{Bodenheimer}}},\ }\href {\doibase 10.1016/j.icarus.2020.114087} {\bibfield
  {journal} {\bibinfo  {journal} {\icarus}\ }\textbf {\bibinfo {volume}
  {355}},\ \bibinfo {eid} {114087} (\bibinfo {year} {2021})},\ \Eprint
  {http://arxiv.org/abs/2009.05575} {arXiv:2009.05575 [astro-ph.EP]}
  \BibitemShut {NoStop}%
\bibitem [{\citenamefont {Freytag}\ \emph {et~al.}(2010)\citenamefont
  {Freytag}, \citenamefont {Allard}, \citenamefont {Ludwig}, \citenamefont
  {Homeier},\ and\ \citenamefont {Steffen}}]{Freytag:2010kn}%
  \BibitemOpen
  \bibfield  {author} {\bibinfo {author} {\bibfnamefont {B.}~\bibnamefont
  {Freytag}}, \bibinfo {author} {\bibfnamefont {F.}~\bibnamefont {Allard}},
  \bibinfo {author} {\bibfnamefont {H.-G.}\ \bibnamefont {Ludwig}}, \bibinfo
  {author} {\bibfnamefont {D.}~\bibnamefont {Homeier}}, \ and\ \bibinfo
  {author} {\bibfnamefont {M.}~\bibnamefont {Steffen}},\ }\href {\doibase
  10.1051/0004-6361/200913354} {\bibfield  {journal} {\bibinfo  {journal}
  {Astron. Astrophys.}\ }\textbf {\bibinfo {volume} {513}},\ \bibinfo {pages}
  {A19} (\bibinfo {year} {2010})},\ \Eprint {http://arxiv.org/abs/1002.3437}
  {arXiv:1002.3437 [astro-ph.SR]} \BibitemShut {NoStop}%
\bibitem [{\citenamefont {Lissauer}\ \emph {et~al.}(2009)\citenamefont
  {Lissauer}, \citenamefont {Hubickyj}, \citenamefont {D'Angelo},\ and\
  \citenamefont {Bodenheimer}}]{Lissauer:2008hn}%
  \BibitemOpen
  \bibfield  {author} {\bibinfo {author} {\bibfnamefont {J.~J.}\ \bibnamefont
  {Lissauer}}, \bibinfo {author} {\bibfnamefont {O.}~\bibnamefont {Hubickyj}},
  \bibinfo {author} {\bibfnamefont {G.}~\bibnamefont {D'Angelo}}, \ and\
  \bibinfo {author} {\bibfnamefont {P.}~\bibnamefont {Bodenheimer}},\ }\href
  {\doibase 10.1016/j.icarus.2008.10.004} {\bibfield  {journal} {\bibinfo
  {journal} {Icarus}\ }\textbf {\bibinfo {volume} {199}},\ \bibinfo {pages}
  {338} (\bibinfo {year} {2009})},\ \Eprint {http://arxiv.org/abs/0810.5186}
  {arXiv:0810.5186 [astro-ph]} \BibitemShut {NoStop}%
\bibitem [{\citenamefont {Jeans}(2009)}]{jeans_2009}%
  \BibitemOpen
  \bibfield  {author} {\bibinfo {author} {\bibfnamefont {J.}~\bibnamefont
  {Jeans}},\ }\href {\doibase 10.1017/CBO9780511694370} {\emph {\bibinfo
  {title} {The Dynamical Theory of Gases}}},\ \bibinfo {edition} {4th}\ ed.,\
  Cambridge Library Collection - Physical Sciences\ (\bibinfo  {publisher}
  {Cambridge University Press},\ \bibinfo {year} {2009})\BibitemShut {NoStop}%
\bibitem [{\citenamefont {Yi{\u{g}}it}(2021)}]{yiugit2021martian}%
  \BibitemOpen
  \bibfield  {author} {\bibinfo {author} {\bibfnamefont {E.}~\bibnamefont
  {Yi{\u{g}}it}},\ }\href@noop {} {\bibfield  {journal} {\bibinfo  {journal}
  {Science}\ }\textbf {\bibinfo {volume} {374}},\ \bibinfo {pages} {1323}
  (\bibinfo {year} {2021})}\BibitemShut {NoStop}%
\bibitem [{\citenamefont {D'Angelo}\ \emph {et~al.}(2021)\citenamefont
  {D'Angelo}, \citenamefont {Weidenschilling}, \citenamefont {Lissauer},\ and\
  \citenamefont {Bodenheimer}}]{d2021growth}%
  \BibitemOpen
  \bibfield  {author} {\bibinfo {author} {\bibfnamefont {G.}~\bibnamefont
  {D'Angelo}}, \bibinfo {author} {\bibfnamefont {S.~J.}\ \bibnamefont
  {Weidenschilling}}, \bibinfo {author} {\bibfnamefont {J.~J.}\ \bibnamefont
  {Lissauer}}, \ and\ \bibinfo {author} {\bibfnamefont {P.}~\bibnamefont
  {Bodenheimer}},\ }\href@noop {} {\bibfield  {journal} {\bibinfo  {journal}
  {Icarus}\ }\textbf {\bibinfo {volume} {355}},\ \bibinfo {pages} {114087}
  (\bibinfo {year} {2021})}\BibitemShut {NoStop}%
\bibitem [{\citenamefont {Leane}\ and\ \citenamefont
  {Smirnov}(2023{\natexlab{a}})}]{RKLandJS}%
  \BibitemOpen
  \bibfield  {author} {\bibinfo {author} {\bibfnamefont {R.}~\bibnamefont
  {Leane}}\ and\ \bibinfo {author} {\bibfnamefont {J.}~\bibnamefont
  {Smirnov}},\ }\href@noop {} {\  (\bibinfo {year} {2023}{\natexlab{a}})},\
  \Eprint {http://arxiv.org/abs/2309.00669} {arXiv:2309.00669 [hep-ph]}
  \BibitemShut {NoStop}%
\bibitem [{\citenamefont {Leane}\ and\ \citenamefont
  {Smirnov}(2023{\natexlab{b}})}]{asteria}%
  \BibitemOpen
  \bibfield  {author} {\bibinfo {author} {\bibfnamefont {R.~K.}\ \bibnamefont
  {Leane}}\ and\ \bibinfo {author} {\bibfnamefont {J.}~\bibnamefont
  {Smirnov}},\ }\href@noop {} {\enquote {\bibinfo {title} {{Asteria: A Package
  for Dark Matter Capture in Celestial Objects}},}\ }\bibinfo {howpublished}
  {\url{https://zenodo.org/record/8150110}} (\bibinfo {year}
  {2023}{\natexlab{b}})\BibitemShut {NoStop}%
\bibitem [{\citenamefont {Digman}\ \emph {et~al.}(2019)\citenamefont {Digman},
  \citenamefont {Cappiello}, \citenamefont {Beacom}, \citenamefont {Hirata},\
  and\ \citenamefont {Peter}}]{Digman:2019wdm}%
  \BibitemOpen
  \bibfield  {author} {\bibinfo {author} {\bibfnamefont {M.~C.}\ \bibnamefont
  {Digman}}, \bibinfo {author} {\bibfnamefont {C.~V.}\ \bibnamefont
  {Cappiello}}, \bibinfo {author} {\bibfnamefont {J.~F.}\ \bibnamefont
  {Beacom}}, \bibinfo {author} {\bibfnamefont {C.~M.}\ \bibnamefont {Hirata}},
  \ and\ \bibinfo {author} {\bibfnamefont {A.~H.~G.}\ \bibnamefont {Peter}},\
  }\href {\doibase 10.1103/PhysRevD.100.063013} {\bibfield  {journal} {\bibinfo
   {journal} {Phys. Rev. D}\ }\textbf {\bibinfo {volume} {100}},\ \bibinfo
  {pages} {063013} (\bibinfo {year} {2019})},\ \Eprint
  {http://arxiv.org/abs/1907.10618} {arXiv:1907.10618 [hep-ph]} \BibitemShut
  {NoStop}%
\bibitem [{\citenamefont {Cappiello}\ \emph {et~al.}(2021)\citenamefont
  {Cappiello}, \citenamefont {Collar},\ and\ \citenamefont
  {Beacom}}]{Cappiello:2020lbk}%
  \BibitemOpen
  \bibfield  {author} {\bibinfo {author} {\bibfnamefont {C.~V.}\ \bibnamefont
  {Cappiello}}, \bibinfo {author} {\bibfnamefont {J.~I.}\ \bibnamefont
  {Collar}}, \ and\ \bibinfo {author} {\bibfnamefont {J.~F.}\ \bibnamefont
  {Beacom}},\ }\href {\doibase 10.1103/PhysRevD.103.023019} {\bibfield
  {journal} {\bibinfo  {journal} {Phys. Rev. D}\ }\textbf {\bibinfo {volume}
  {103}},\ \bibinfo {pages} {023019} (\bibinfo {year} {2021})},\ \Eprint
  {http://arxiv.org/abs/2008.10646} {arXiv:2008.10646 [hep-ex]} \BibitemShut
  {NoStop}%
\bibitem [{\citenamefont {Mack}\ \emph {et~al.}(2007)\citenamefont {Mack},
  \citenamefont {Beacom},\ and\ \citenamefont {Bertone}}]{Mack:2007xj}%
  \BibitemOpen
  \bibfield  {author} {\bibinfo {author} {\bibfnamefont {G.~D.}\ \bibnamefont
  {Mack}}, \bibinfo {author} {\bibfnamefont {J.~F.}\ \bibnamefont {Beacom}}, \
  and\ \bibinfo {author} {\bibfnamefont {G.}~\bibnamefont {Bertone}},\ }\href
  {\doibase 10.1103/PhysRevD.76.043523} {\bibfield  {journal} {\bibinfo
  {journal} {Phys. Rev. D}\ }\textbf {\bibinfo {volume} {76}},\ \bibinfo
  {pages} {043523} (\bibinfo {year} {2007})},\ \Eprint
  {http://arxiv.org/abs/0705.4298} {arXiv:0705.4298 [astro-ph]} \BibitemShut
  {NoStop}%
\bibitem [{\citenamefont {Acevedo}\ \emph {et~al.}(2023)\citenamefont
  {Acevedo}, \citenamefont {Leane},\ and\ \citenamefont
  {Smirnov}}]{Acevedo:2023owd}%
  \BibitemOpen
  \bibfield  {author} {\bibinfo {author} {\bibfnamefont {J.~F.}\ \bibnamefont
  {Acevedo}}, \bibinfo {author} {\bibfnamefont {R.~K.}\ \bibnamefont {Leane}},
  \ and\ \bibinfo {author} {\bibfnamefont {J.}~\bibnamefont {Smirnov}},\
  }\href@noop {} {\  (\bibinfo {year} {2023})},\ \Eprint
  {http://arxiv.org/abs/2303.01516} {arXiv:2303.01516 [hep-ph]} \BibitemShut
  {NoStop}%
\bibitem [{\citenamefont {Boddy}\ and\ \citenamefont
  {Gluscevic}(2018)}]{Boddy:2018kfv}%
  \BibitemOpen
  \bibfield  {author} {\bibinfo {author} {\bibfnamefont {K.~K.}\ \bibnamefont
  {Boddy}}\ and\ \bibinfo {author} {\bibfnamefont {V.}~\bibnamefont
  {Gluscevic}},\ }\href {\doibase 10.1103/PhysRevD.98.083510} {\bibfield
  {journal} {\bibinfo  {journal} {Phys. Rev. D}\ }\textbf {\bibinfo {volume}
  {98}},\ \bibinfo {pages} {083510} (\bibinfo {year} {2018})},\ \Eprint
  {http://arxiv.org/abs/1801.08609} {arXiv:1801.08609 [astro-ph.CO]}
  \BibitemShut {NoStop}%
\bibitem [{\citenamefont {Nadler}\ \emph {et~al.}(2019)\citenamefont {Nadler},
  \citenamefont {Gluscevic}, \citenamefont {Boddy},\ and\ \citenamefont
  {Wechsler}}]{Nadler:2019zrb}%
  \BibitemOpen
  \bibfield  {author} {\bibinfo {author} {\bibfnamefont {E.~O.}\ \bibnamefont
  {Nadler}}, \bibinfo {author} {\bibfnamefont {V.}~\bibnamefont {Gluscevic}},
  \bibinfo {author} {\bibfnamefont {K.~K.}\ \bibnamefont {Boddy}}, \ and\
  \bibinfo {author} {\bibfnamefont {R.~H.}\ \bibnamefont {Wechsler}},\ }\href
  {\doibase 10.3847/2041-8213/ab1eb2} {\bibfield  {journal} {\bibinfo
  {journal} {Astrophys. J. Lett.}\ }\textbf {\bibinfo {volume} {878}},\
  \bibinfo {pages} {32} (\bibinfo {year} {2019})},\ \bibinfo {note} {[Erratum:
  Astrophys.J.Lett. 897, L46 (2020), Erratum: Astrophys.J. 897, L46 (2020)]},\
  \Eprint {http://arxiv.org/abs/1904.10000} {arXiv:1904.10000 [astro-ph.CO]}
  \BibitemShut {NoStop}%
\bibitem [{\citenamefont {Sahu}\ \emph {et~al.}(2006)\citenamefont {Sahu} \emph
  {et~al.}}]{Sahu:2006ex}%
  \BibitemOpen
  \bibfield  {author} {\bibinfo {author} {\bibfnamefont {K.~C.}\ \bibnamefont
  {Sahu}} \emph {et~al.},\ }\href {\doibase 10.1038/nature05158} {\bibfield
  {journal} {\bibinfo  {journal} {Nature}\ }\textbf {\bibinfo {volume} {443}},\
  \bibinfo {pages} {534} (\bibinfo {year} {2006})},\ \Eprint
  {http://arxiv.org/abs/astro-ph/0610098} {arXiv:astro-ph/0610098} \BibitemShut
  {NoStop}%
\bibitem [{\citenamefont {Janczak}\ \emph {et~al.}(2010)\citenamefont {Janczak}
  \emph {et~al.}}]{MOA:2009dkq}%
  \BibitemOpen
  \bibfield  {author} {\bibinfo {author} {\bibfnamefont {J.}~\bibnamefont
  {Janczak}} \emph {et~al.} (\bibinfo {collaboration} {MOA, muFUN, MiNDSTEp
  Consortium, PLANET}),\ }\href {\doibase 10.1088/0004-637X/711/2/731}
  {\bibfield  {journal} {\bibinfo  {journal} {Astrophys. J.}\ }\textbf
  {\bibinfo {volume} {711}},\ \bibinfo {pages} {731} (\bibinfo {year}
  {2010})},\ \Eprint {http://arxiv.org/abs/0908.0529} {arXiv:0908.0529
  [astro-ph.EP]} \BibitemShut {NoStop}%
\bibitem [{\citenamefont {Koshimoto}\ \emph {et~al.}(2021)\citenamefont
  {Koshimoto}, \citenamefont {Bennett}, \citenamefont {Suzuki},\ and\
  \citenamefont {Bond}}]{koshimoto2021no}%
  \BibitemOpen
  \bibfield  {author} {\bibinfo {author} {\bibfnamefont {N.}~\bibnamefont
  {Koshimoto}}, \bibinfo {author} {\bibfnamefont {D.~P.}\ \bibnamefont
  {Bennett}}, \bibinfo {author} {\bibfnamefont {D.}~\bibnamefont {Suzuki}}, \
  and\ \bibinfo {author} {\bibfnamefont {I.~A.}\ \bibnamefont {Bond}},\
  }\href@noop {} {\bibfield  {journal} {\bibinfo  {journal} {The Astrophysical
  Journal Letters}\ }\textbf {\bibinfo {volume} {918}},\ \bibinfo {pages} {L8}
  (\bibinfo {year} {2021})}\BibitemShut {NoStop}%
\bibitem [{\citenamefont {Angloher}\ \emph {et~al.}(2023)\citenamefont
  {Angloher} \emph {et~al.}}]{CRESST:2022lqw}%
  \BibitemOpen
  \bibfield  {author} {\bibinfo {author} {\bibfnamefont {G.}~\bibnamefont
  {Angloher}} \emph {et~al.} (\bibinfo {collaboration} {CRESST}),\ }\href
  {\doibase 10.1103/PhysRevD.107.122003} {\bibfield  {journal} {\bibinfo
  {journal} {Phys. Rev. D}\ }\textbf {\bibinfo {volume} {107}},\ \bibinfo
  {pages} {122003} (\bibinfo {year} {2023})},\ \Eprint
  {http://arxiv.org/abs/2212.12513} {arXiv:2212.12513 [astro-ph.CO]}
  \BibitemShut {NoStop}%
\bibitem [{\citenamefont {Angloher}\ \emph {et~al.}(2022)\citenamefont
  {Angloher} \emph {et~al.}}]{CRESST:2022dtl}%
  \BibitemOpen
  \bibfield  {author} {\bibinfo {author} {\bibfnamefont {G.}~\bibnamefont
  {Angloher}} \emph {et~al.} (\bibinfo {collaboration} {CRESST}),\ }\href
  {\doibase 10.1103/PhysRevD.106.092008} {\bibfield  {journal} {\bibinfo
  {journal} {Phys. Rev. D}\ }\textbf {\bibinfo {volume} {106}},\ \bibinfo
  {pages} {092008} (\bibinfo {year} {2022})},\ \Eprint
  {http://arxiv.org/abs/2207.07640} {arXiv:2207.07640 [astro-ph.CO]}
  \BibitemShut {NoStop}%
\bibitem [{\citenamefont {Akerib}\ \emph {et~al.}(2017)\citenamefont {Akerib}
  \emph {et~al.}}]{LUX:2017ree}%
  \BibitemOpen
  \bibfield  {author} {\bibinfo {author} {\bibfnamefont {D.~S.}\ \bibnamefont
  {Akerib}} \emph {et~al.} (\bibinfo {collaboration} {LUX}),\ }\href {\doibase
  10.1103/PhysRevLett.118.251302} {\bibfield  {journal} {\bibinfo  {journal}
  {Phys. Rev. Lett.}\ }\textbf {\bibinfo {volume} {118}},\ \bibinfo {pages}
  {251302} (\bibinfo {year} {2017})},\ \Eprint
  {http://arxiv.org/abs/1705.03380} {arXiv:1705.03380 [astro-ph.CO]}
  \BibitemShut {NoStop}%
\bibitem [{\citenamefont {Aprile}\ \emph {et~al.}(2023)\citenamefont {Aprile}
  \emph {et~al.}}]{XENONCollaboration:2023orw}%
  \BibitemOpen
  \bibfield  {author} {\bibinfo {author} {\bibfnamefont {E.}~\bibnamefont
  {Aprile}} \emph {et~al.} (\bibinfo {collaboration} {(XENON
  Collaboration)\textdagger{}\textdagger{}, XENON}),\ }\href {\doibase
  10.1103/PhysRevLett.131.041003} {\bibfield  {journal} {\bibinfo  {journal}
  {Phys. Rev. Lett.}\ }\textbf {\bibinfo {volume} {131}},\ \bibinfo {pages}
  {041003} (\bibinfo {year} {2023})},\ \Eprint
  {http://arxiv.org/abs/2303.14729} {arXiv:2303.14729 [hep-ex]} \BibitemShut
  {NoStop}%
\bibitem [{\citenamefont {Erickcek}\ \emph {et~al.}(2007)\citenamefont
  {Erickcek}, \citenamefont {Steinhardt}, \citenamefont {McCammon},\ and\
  \citenamefont {McGuire}}]{Erickcek:2007jv}%
  \BibitemOpen
  \bibfield  {author} {\bibinfo {author} {\bibfnamefont {A.~L.}\ \bibnamefont
  {Erickcek}}, \bibinfo {author} {\bibfnamefont {P.~J.}\ \bibnamefont
  {Steinhardt}}, \bibinfo {author} {\bibfnamefont {D.}~\bibnamefont
  {McCammon}}, \ and\ \bibinfo {author} {\bibfnamefont {P.~C.}\ \bibnamefont
  {McGuire}},\ }\href {\doibase 10.1103/PhysRevD.76.042007} {\bibfield
  {journal} {\bibinfo  {journal} {Phys. Rev.}\ }\textbf {\bibinfo {volume}
  {D76}},\ \bibinfo {pages} {042007} (\bibinfo {year} {2007})},\ \Eprint
  {http://arxiv.org/abs/0704.0794} {arXiv:0704.0794 [astro-ph]} \BibitemShut
  {NoStop}%
\bibitem [{\citenamefont {Mahdawi}\ and\ \citenamefont
  {Farrar}(2018)}]{Mahdawi:2018euy}%
  \BibitemOpen
  \bibfield  {author} {\bibinfo {author} {\bibfnamefont {M.~S.}\ \bibnamefont
  {Mahdawi}}\ and\ \bibinfo {author} {\bibfnamefont {G.~R.}\ \bibnamefont
  {Farrar}},\ }\href {\doibase 10.1088/1475-7516/2018/10/007} {\bibfield
  {journal} {\bibinfo  {journal} {JCAP}\ }\textbf {\bibinfo {volume} {10}},\
  \bibinfo {pages} {007} (\bibinfo {year} {2018})},\ \Eprint
  {http://arxiv.org/abs/1804.03073} {arXiv:1804.03073 [hep-ph]} \BibitemShut
  {NoStop}%
\bibitem [{\citenamefont {Albakry}\ \emph {et~al.}(2023)\citenamefont {Albakry}
  \emph {et~al.}}]{SuperCDMS:2023sql}%
  \BibitemOpen
  \bibfield  {author} {\bibinfo {author} {\bibfnamefont {M.~F.}\ \bibnamefont
  {Albakry}} \emph {et~al.} (\bibinfo {collaboration} {SuperCDMS}),\ }\href
  {\doibase 10.1103/PhysRevD.107.112013} {\bibfield  {journal} {\bibinfo
  {journal} {Phys. Rev. D}\ }\textbf {\bibinfo {volume} {107}},\ \bibinfo
  {pages} {112013} (\bibinfo {year} {2023})},\ \Eprint
  {http://arxiv.org/abs/2302.09115} {arXiv:2302.09115 [hep-ex]} \BibitemShut
  {NoStop}%
\bibitem [{\citenamefont {Cox}\ \emph {et~al.}(2023)\citenamefont {Cox},
  \citenamefont {Dolan}, \citenamefont {McCabe},\ and\ \citenamefont
  {Quiney}}]{Cox:2022ekg}%
  \BibitemOpen
  \bibfield  {author} {\bibinfo {author} {\bibfnamefont {P.}~\bibnamefont
  {Cox}}, \bibinfo {author} {\bibfnamefont {M.~J.}\ \bibnamefont {Dolan}},
  \bibinfo {author} {\bibfnamefont {C.}~\bibnamefont {McCabe}}, \ and\ \bibinfo
  {author} {\bibfnamefont {H.~M.}\ \bibnamefont {Quiney}},\ }\href {\doibase
  10.1103/PhysRevD.107.035032} {\bibfield  {journal} {\bibinfo  {journal}
  {Phys. Rev. D}\ }\textbf {\bibinfo {volume} {107}},\ \bibinfo {pages}
  {035032} (\bibinfo {year} {2023})},\ \Eprint
  {http://arxiv.org/abs/2208.12222} {arXiv:2208.12222 [hep-ph]} \BibitemShut
  {NoStop}%
\bibitem [{\citenamefont {Bramante}\ \emph {et~al.}(2020)\citenamefont
  {Bramante}, \citenamefont {Buchanan}, \citenamefont {Goodman},\ and\
  \citenamefont {Lodhi}}]{Bramante:2019fhi}%
  \BibitemOpen
  \bibfield  {author} {\bibinfo {author} {\bibfnamefont {J.}~\bibnamefont
  {Bramante}}, \bibinfo {author} {\bibfnamefont {A.}~\bibnamefont {Buchanan}},
  \bibinfo {author} {\bibfnamefont {A.}~\bibnamefont {Goodman}}, \ and\
  \bibinfo {author} {\bibfnamefont {E.}~\bibnamefont {Lodhi}},\ }\href
  {\doibase 10.1103/PhysRevD.101.043001} {\bibfield  {journal} {\bibinfo
  {journal} {Phys. Rev. D}\ }\textbf {\bibinfo {volume} {101}},\ \bibinfo
  {pages} {043001} (\bibinfo {year} {2020})},\ \Eprint
  {http://arxiv.org/abs/1909.11683} {arXiv:1909.11683 [hep-ph]} \BibitemShut
  {NoStop}%
\bibitem [{\citenamefont {Cappiello}(2023)}]{Cappiello:2023hza}%
  \BibitemOpen
  \bibfield  {author} {\bibinfo {author} {\bibfnamefont {C.~V.}\ \bibnamefont
  {Cappiello}},\ }\href {\doibase 10.1103/PhysRevLett.130.221001} {\bibfield
  {journal} {\bibinfo  {journal} {Phys. Rev. Lett.}\ }\textbf {\bibinfo
  {volume} {130}},\ \bibinfo {pages} {221001} (\bibinfo {year} {2023})},\
  \Eprint {http://arxiv.org/abs/2301.07728} {arXiv:2301.07728 [hep-ph]}
  \BibitemShut {NoStop}%
\bibitem [{\citenamefont {Kavanagh}(2018)}]{Kavanagh:2017cru}%
  \BibitemOpen
  \bibfield  {author} {\bibinfo {author} {\bibfnamefont {B.~J.}\ \bibnamefont
  {Kavanagh}},\ }\href {\doibase 10.1103/PhysRevD.97.123013} {\bibfield
  {journal} {\bibinfo  {journal} {Phys. Rev. D}\ }\textbf {\bibinfo {volume}
  {97}},\ \bibinfo {pages} {123013} (\bibinfo {year} {2018})},\ \Eprint
  {http://arxiv.org/abs/1712.04901} {arXiv:1712.04901 [hep-ph]} \BibitemShut
  {NoStop}%
\bibitem [{\citenamefont {Leane}\ and\ \citenamefont
  {Smirnov}(2022)}]{Leane:2022hkk}%
  \BibitemOpen
  \bibfield  {author} {\bibinfo {author} {\bibfnamefont {R.~K.}\ \bibnamefont
  {Leane}}\ and\ \bibinfo {author} {\bibfnamefont {J.}~\bibnamefont
  {Smirnov}},\ }\href@noop {} {\  (\bibinfo {year} {2022})},\ \Eprint
  {http://arxiv.org/abs/2209.09834} {arXiv:2209.09834 [hep-ph]} \BibitemShut
  {NoStop}%
\bibitem [{\citenamefont {Elor}\ \emph {et~al.}(2023)\citenamefont {Elor},
  \citenamefont {McGehee},\ and\ \citenamefont {Pierce}}]{Elor:2021swj}%
  \BibitemOpen
  \bibfield  {author} {\bibinfo {author} {\bibfnamefont {G.}~\bibnamefont
  {Elor}}, \bibinfo {author} {\bibfnamefont {R.}~\bibnamefont {McGehee}}, \
  and\ \bibinfo {author} {\bibfnamefont {A.}~\bibnamefont {Pierce}},\ }\href
  {\doibase 10.1103/PhysRevLett.130.031803} {\bibfield  {journal} {\bibinfo
  {journal} {Phys. Rev. Lett.}\ }\textbf {\bibinfo {volume} {130}},\ \bibinfo
  {pages} {031803} (\bibinfo {year} {2023})},\ \Eprint
  {http://arxiv.org/abs/2112.03920} {arXiv:2112.03920 [hep-ph]} \BibitemShut
  {NoStop}%
\bibitem [{\citenamefont {{Gould}}(1987)}]{1987ApJ...321..560G}%
  \BibitemOpen
  \bibfield  {author} {\bibinfo {author} {\bibfnamefont {A.}~\bibnamefont
  {{Gould}}},\ }\href {\doibase 10.1086/165652} {\bibfield  {journal} {\bibinfo
   {journal} {\apj}\ }\textbf {\bibinfo {volume} {321}},\ \bibinfo {pages}
  {560} (\bibinfo {year} {1987})}\BibitemShut {NoStop}%
\bibitem [{\citenamefont {{Gould}}(1990)}]{1990ApJ...356..302G}%
  \BibitemOpen
  \bibfield  {author} {\bibinfo {author} {\bibfnamefont {A.}~\bibnamefont
  {{Gould}}},\ }\href {\doibase 10.1086/168840} {\bibfield  {journal} {\bibinfo
   {journal} {\apj}\ }\textbf {\bibinfo {volume} {356}},\ \bibinfo {pages}
  {302} (\bibinfo {year} {1990})}\BibitemShut {NoStop}%
\end{thebibliography}%
